\documentclass[journal,10pt]{IEEEtran}
\usepackage{amsmath,amssymb}
\usepackage{graphicx}
\graphicspath{{./images_for_presentation}}
\usepackage{booktabs}
\usepackage{caption}
\usepackage{subcaption}
\usepackage{verbatim}
\usepackage{hyperref}
\usepackage{cite}
\usepackage{tikz}

\title{Alternative Shapes of Modulation Schemes\\
Detailed Exposition and Simulation Methodology}
\author{Nipun Agarwal\\
Email: nipuna@kth.se}

\hypersetup{
    colorlinks=true,
    linkcolor=red,
    citecolor=green,
    filecolor=magenta,      
    urlcolor=blue,
    pdftitle={Overleaf Example},
    pdfpagemode=FullScreen,
    }

\begin{document}
\maketitle

\begin{abstract}
Modulation constellation design remains a fundamental yet evolving problem in digital communication systems, particularly as modern networks face stringent requirements on spectral efficiency, robustness, and energy consumption. While classical modulation schemes such as phase-shift keying (PSK) and quadrature amplitude modulation (QAM) are analytically tractable and widely deployed, their optimality degrades under realistic channel conditions and transmitter hardware constraints, especially nonlinear power amplification.

This paper presents a comprehensive and unified investigation of modulation constellation design from geometric, probabilistic, optimization-based, and machine learning perspectives, with explicit emphasis on symbol error rate (SER), fading robustness, peak-to-average power ratio (PAPR), and energy efficiency. A diverse set of modulation schemes—including classical constellations, lattice-based and asymmetric designs, probabilistically shaped constellations, Golden Angle Modulation, heuristic optimization–based constellations, and machine learning–assisted designs—are evaluated under additive white Gaussian noise and Rayleigh fading channels using a large-scale Monte Carlo framework.

Beyond conventional SER analysis, the study incorporates transmitter energy consumption through PAPR-aware and power amplifier efficiency models, revealing that SER-optimal constellations are not necessarily energy-optimal. Numerical results demonstrate that modest degradations in error performance can yield substantial energy savings, particularly for low-PAPR and geometry-constrained designs. Furthermore, machine learning–assisted constellation optimization is shown to provide a flexible framework capable of jointly optimizing reliability, robustness, and energy efficiency by embedding channel statistics and hardware models directly into the learning objective.

The paper culminates in a set of practical design guidelines that map operating regimes, channel conditions, and energy constraints to appropriate modulation strategies. Collectively, the results establish modulation design as a multi-objective optimization problem and provide a principled foundation for the development of energy-aware and intelligent modulation schemes in next-generation wireless communication systems.

\end{abstract}


\section{Introduction}

Digital modulation lies at the core of modern communication systems, directly governing spectral efficiency, power efficiency, robustness to noise and fading, and hardware complexity. For several decades, square quadrature amplitude modulation (QAM) and phase-shift keying (PSK) constellations have dominated both wireless and wireline standards due to their structural regularity, ease of Gray labeling, and compatibility with linear receivers and hardware implementations \cite{Forney1984,Proakis2008}. However, these classical constellations were largely designed under assumptions of ideal linear channels and simplified transmitter architectures, assumptions that are increasingly violated in contemporary communication systems.

The exponential growth of mobile traffic, dense network deployments, and stringent energy efficiency requirements have exposed the fundamental limitations of conventional modulation schemes. Square QAM, in particular, suffers from a well-known shaping loss of approximately 1.53~dB relative to the theoretical capacity of the additive white Gaussian noise (AWGN) channel \cite{Kschischang1993}. Moreover, its relatively high peak-to-average power ratio (PAPR) necessitates substantial power amplifier (PA) back-off, which significantly reduces transmitter efficiency and increases operational energy consumption \cite{Paul2015,Lorincz2019}. These shortcomings become even more pronounced in systems operating near saturation, such as millimetre-wave transmitters, massive multiple-input multiple-output (MIMO) arrays, and battery-powered user equipment.

In parallel, advances in information theory have long established that Gaussian-distributed channel inputs are capacity-achieving for AWGN channels \cite{Cover2006}. This insight has motivated extensive research into non-uniform signaling and constellation shaping techniques that more closely approximate Gaussian distributions while retaining finite alphabet structures suitable for practical implementation \cite{Betts1994,Bocherer2015}. Early works on optimal nonuniform signaling demonstrated significant potential gains in both power efficiency and achievable information rates, but practical realization remained challenging due to mapping complexity and sensitivity to implementation constraints \cite{Kschischang1993}.

More recently, probabilistic shaping and geometric shaping have emerged as powerful and complementary approaches for constellation optimization. Probabilistic shaping modifies symbol occurrence probabilities—often following a Maxwell–Boltzmann distribution—while maintaining a fixed geometric layout \cite{Bocherer2015,Zhang2017}. Geometric shaping, by contrast, directly alters constellation point locations to improve minimum Euclidean distance, reduce PAPR, or adapt to channel characteristics \cite{Barrueco2021}. Both approaches have demonstrated measurable gains over conventional QAM in optical and wireless systems, particularly when combined with advanced coding schemes.

Beyond analytically designed constellations, optimization-based techniques have gained significant traction. Evolutionary algorithms such as genetic algorithms (GA) and particle swarm optimization (PSO) enable the automated discovery of irregular constellation geometries tailored to specific objective functions, including symbol error rate (SER), mutual information, or hardware-aware metrics \cite{Eberhart1995,Ishibashi2014}. These methods are especially attractive for scenarios where closed-form solutions are intractable, such as fading channels, nonlinear hardware models, or multi-objective optimization involving both performance and energy efficiency.

A particularly elegant class of geometrically shaped constellations is Golden Angle Modulation (GAM), where constellation points are placed along a spiral using the golden angle increment \cite{Larsson2018,Sharma2020}. GAM variants, including Disc-GAM and Bell-GAM, have been shown to approximate Gaussian-like distributions while exhibiting reduced PAPR and strong robustness to nonlinear distortion \cite{Ibragimov2017,Mheich2019}. Their simple parametric structure and near-capacity performance make them promising candidates for future adaptive modulation schemes.

Despite the breadth of existing work, several important gaps remain. First, many studies focus on isolated constellation families without providing a unified, reproducible comparison across classical, geometric, probabilistic, and ML-optimized designs. Second, performance evaluations often emphasize SER or mutual information alone, neglecting the critical role of PAPR and its downstream impact on PA efficiency and energy consumption. Third, sustainability considerations—such as network-level energy savings and carbon footprint reduction—are rarely quantified in a systematic manner.

\subsection{Contributions and Hypotheses}

This paper addresses these gaps through a comprehensive and methodical investigation of alternative modulation schemes. The main contributions are summarised as follows:

\begin{itemize}
\item A unified mathematical and algorithmic framework for fourteen modulation constellations, spanning classical, lattice-based, optimization-driven, probabilistically shaped, and spiral-based designs.
\item A fully reproducible Monte Carlo simulation methodology for SER evaluation over AWGN and Rayleigh fading channels, including statistical confidence analysis.
\item An explicit energy-consumption model linking constellation PAPR to efficiency, and network-level sustainability metrics.
\item An exploration of machine-learning-assisted constellation design and demapping techniques, highlighting their practical benefits and limitations.
\end{itemize}

The investigation is guided by the following hypotheses:

\begin{itemize}
\item[\textbf{H1}] Properly shaped non-uniform constellations can achieve lower SER than square QAM at the same average symbol energy.
\item[\textbf{H2}] Constellations with reduced PAPR yield tangible energy-efficiency gains at the transmitter due to improved PA efficiency.
\item[\textbf{H3}] Optimization-based and ML-assisted constellation design can outperform analytically designed constellations under practical constraints.
\item[\textbf{H4}] Rayleigh fading will penalize uniform constellations more severely than optimized
asymmetric designs.
\end{itemize}

The remainder of this paper is organized as follows. Section~II provides a detailed mathematical description of all modulation schemes under study. Section~III presents the simulation methodology and statistical analysis framework. Machine-learning approaches are discussed in Section~IV, while Section~V quantifies energy consumption and sustainability implications. Numerical results and discussions follow, concluding with insights and directions for future research.

\section{Detailed Modulation Schemes}
\label{sec:modulation}

This section provides an in-depth and systematic treatment of the modulation constellations investigated in this work. Rather than limiting the discussion to geometric descriptions alone, the section establishes a rigorous analytical foundation, introduces key performance metrics, and highlights the fundamental trade-offs that govern modulation design under practical channel and hardware constraints. Rather than evaluating each modulation scheme in isolation, this section emphasizes how different geometric and probabilistic design philosophies influence reliability, robustness, and energy efficiency. Detailed performance implications are quantified later in Sections~\ref{sec:system_model} and \ref{sec:numerical_results}.

\subsection{Mathematical Preliminaries and Design Metrics}
\label{sec:preliminaries}

The performance of any digital modulation scheme is fundamentally governed by its geometric, probabilistic, and statistical properties in signal space. Before presenting specific constellation designs, this subsection establishes a rigorous mathematical framework and defines the principal metrics used throughout this work. These preliminaries ensure that all modulation schemes are evaluated under a unified and fair analytical basis.

\subsubsection{Complex Baseband Signal Model}

Modern digital communication systems are commonly represented using a complex baseband equivalent model. Let the transmitted symbol sequence be denoted by
\begin{equation}
x[n] \in \mathcal{S} = \{s_1, s_2, \ldots, s_M\},
\end{equation}
where $\mathcal{S}$ is an $M$-ary constellation defined over the complex plane $\mathbb{C}$. The received signal over an additive white Gaussian noise (AWGN) channel is given by
\begin{equation}
y[n] = x[n] + w[n],
\end{equation}
where $w[n] \sim \mathcal{CN}(0, N_0)$ is a circularly symmetric complex Gaussian noise process with two-sided power spectral density $N_0/2$ per real dimension \cite{Proakis2008}.

For flat fading channels, the model generalizes to
\begin{equation}
y[n] = h[n] x[n] + w[n],
\end{equation}
where $h[n]$ is a complex channel coefficient. In Rayleigh fading, $h[n] \sim \mathcal{CN}(0,1)$. Unless otherwise stated, perfect channel state information (CSI) at the receiver is assumed in order to isolate the intrinsic effects of constellation geometry.

\subsubsection{Signal Space Representation}

Each constellation point $s_k$ can be represented as a two-dimensional real vector
\begin{equation}
s_k = s_{k,I} + j s_{k,Q}
\quad \leftrightarrow \quad
\mathbf{s}_k =
\begin{bmatrix}
s_{k,I} \\
s_{k,Q}
\end{bmatrix}
\in \mathbb{R}^2.
\end{equation}
This interpretation allows modulation design to be studied using tools from Euclidean geometry and lattice theory. The squared Euclidean distance between two constellation points is
\begin{equation}
d_{ij}^2 = |s_i - s_j|^2
= (s_{i,I} - s_{j,I})^2 + (s_{i,Q} - s_{j,Q})^2.
\end{equation}

\subsubsection{Energy Normalization}

To enable meaningful comparisons across different modulation schemes, all constellations are normalized to unit average symbol energy. The average symbol energy is defined as
\begin{equation}
E_s = \mathbb{E}[|X|^2] = \frac{1}{M} \sum_{k=1}^{M} |s_k|^2.
\end{equation}
Normalization is performed via
\begin{equation}
\tilde{s}_k = \frac{s_k}{\sqrt{E_s}},
\end{equation}
ensuring $\mathbb{E}[|\tilde{X}|^2] = 1$. This step is essential for fair SER and PAPR comparisons.

\subsubsection{Signal-to-Noise Ratio Definitions}

The signal-to-noise ratio (SNR) per symbol is defined as
\begin{equation}
\gamma_s = \frac{E_s}{N_0}.
\end{equation}
For an $M$-ary modulation transmitting $k = \log_2 M$ bits per symbol, the SNR per bit is
\begin{equation}
\gamma_b = \frac{E_b}{N_0} = \frac{E_s}{k N_0}.
\end{equation}
Throughout this paper, results are primarily expressed in terms of $\gamma_s$, as it enables modulation-independent comparisons.

\subsubsection{Minimum Euclidean Distance}

The minimum Euclidean distance between constellation points is defined as
\begin{equation}
d_{\min} = \min_{i \neq j} |s_i - s_j|.
\end{equation}
For AWGN channels and maximum-likelihood detection, $d_{\min}$ is the dominant factor governing symbol error rate at moderate-to-high SNR \cite{Forney1984}.

\subsubsection{Distance Spectrum and Union Bound}

The full distance spectrum captures all pairwise distances and their multiplicities. The union bound on SER can be written as
\begin{equation}
P_s \leq \frac{1}{M}
\sum_{i=1}^{M}
\sum_{j \neq i}
Q\left(
\sqrt{\frac{|s_i - s_j|^2}{2N_0}}
\right),
\end{equation}
where $Q(\cdot)$ denotes the Gaussian $Q$-function. While often loose at low SNR, this bound provides valuable insight into high-SNR behavior and constellation robustness.

\subsubsection{Peak-to-Average Power Ratio}

The peak-to-average power ratio (PAPR) is defined as
\begin{equation}
\mathrm{PAPR} =
\frac{\max_k |s_k|^2}{\mathbb{E}[|X|^2]}.
\end{equation}
High PAPR forces power amplifiers to operate with significant back-off, drastically reducing efficiency and increasing energy consumption \cite{Lorincz2019,Paul2015}.

\subsubsection{Mutual Information Perspective}

From an information-theoretic viewpoint, constellation efficiency is measured through mutual information
\begin{equation}
I(X;Y) =
\mathbb{E}\left[
\log_2 \frac{p(Y|X)}{p(Y)}
\right].
\end{equation}
Uniform QAM inputs incur a shaping loss relative to Gaussian signaling, motivating probabilistic and geometric shaping techniques \cite{Cover2006,Bocherer2015}.

\subsubsection{Design Trade-offs and Scope}

Constellation design involves balancing competing objectives, including error performance, spectral efficiency, PAPR, implementation complexity, and robustness to non-ideal hardware. These trade-offs motivate the diverse constellation families studied in the remainder of this section.

\subsection{Classical Modulation Schemes}
\label{sec:classical}

Classical modulation schemes form the foundation of digital communication systems and serve as essential benchmarks for evaluating more advanced constellation designs. Despite their long history and widespread adoption, these schemes exhibit inherent limitations in terms of shaping loss, peak-to-average power ratio (PAPR), and adaptability to modern hardware constraints. This subsection provides a detailed analytical treatment of binary and multilevel classical modulations, highlighting both their strengths and fundamental shortcomings.

\subsubsection{Binary Phase-Shift Keying (BPSK)}

Binary phase-shift keying (BPSK) is the simplest digital modulation scheme, consisting of two antipodal constellation points:
\begin{equation}
\mathcal{S}_{\text{BPSK}} = \{+1, -1\}.
\end{equation}
After normalization, the constellation lies on the real axis with unit energy. The minimum Euclidean distance is
\begin{equation}
d_{\min} = 2,
\end{equation}
which is maximal for a binary constellation under an average energy constraint.

Under AWGN and coherent detection, the symbol error rate (SER) of BPSK is given by
\begin{equation}
P_s = Q\left(\sqrt{2\gamma_s}\right),
\end{equation}
which is identical to the bit error rate (BER) due to the one-to-one symbol-to-bit mapping \cite{Proakis2008}. BPSK achieves optimal power efficiency among all binary constellations and exhibits constant envelope properties, resulting in minimal sensitivity to nonlinear amplification.

However, its spectral efficiency is limited to one bit per symbol, rendering it unsuitable for high-throughput applications.

\subsubsection{Quadrature Phase-Shift Keying (QPSK)}

Quadrature phase-shift keying (QPSK) extends BPSK by transmitting two bits per symbol using four equidistant phase points:
\begin{equation}
s_k = \frac{1}{\sqrt{2}} \left( \pm 1 \pm j \right).
\end{equation}
QPSK can be interpreted as two orthogonal BPSK modulations on the in-phase and quadrature components.

The minimum distance of QPSK is
\begin{equation}
d_{\min} = \sqrt{2},
\end{equation}
and its SER under AWGN is
\begin{equation}
P_s = 2Q\left(\sqrt{\gamma_s}\right) - Q^2\left(\sqrt{\gamma_s}\right).
\end{equation}

QPSK retains constant envelope properties and exhibits excellent robustness to nonlinear distortion, making it widely used in satellite, cellular control channels, and low-SNR regimes \cite{Proakis2008}. However, like BPSK, its spectral efficiency remains limited.

\subsubsection{M-ary Phase-Shift Keying (M-PSK)}

In $M$-PSK, constellation points are uniformly distributed on the unit circle:
\begin{equation}
s_k = e^{j\frac{2\pi k}{M}}, \quad k = 0,1,\dots,M-1.
\end{equation}
The minimum Euclidean distance is
\begin{equation}
d_{\min} = 2\sin\left(\frac{\pi}{M}\right),
\end{equation}
which decreases rapidly as $M$ increases.

An approximate SER expression for coherent detection over AWGN is
\begin{equation}
P_s \approx 2Q\left(\sqrt{2\gamma_s}\sin\left(\frac{\pi}{M}\right)\right),
\end{equation}
valid for moderate-to-high SNR \cite{Proakis2008}.

While $M$-PSK maintains constant envelope properties and thus low PAPR, its rapidly shrinking minimum distance makes it highly sensitive to noise at large constellation orders. Consequently, high-order PSK is rarely used in practice, except in scenarios where constant-envelope transmission is mandatory.
 
\subsubsection{Differential Phase-Shift Keying}

Differential PSK (DPSK) encodes information in the phase difference between successive symbols, eliminating the need for explicit carrier phase recovery. Although DPSK simplifies receiver design, it incurs an SNR penalty of approximately 3~dB relative to coherent PSK due to noise accumulation \cite{Proakis2008}. For this reason, DPSK is primarily used in low-complexity or rapidly time-varying channels.

\subsubsection{Square Quadrature Amplitude Modulation (QAM)}

Square quadrature amplitude modulation (QAM) combines amplitude and phase variations to achieve high spectral efficiency. An $M$-QAM constellation consists of $M = L^2$ points arranged on a square lattice, where
\begin{equation}
\mathcal{A} = \{\pm 1, \pm 3, \dots, \pm (L-1)\}.
\end{equation}
Each symbol is formed as
\begin{equation}
s = a_I + j a_Q, \quad a_I, a_Q \in \mathcal{A}.
\end{equation}

After normalization, square QAM achieves superior spectral efficiency compared to PSK for large $M$. The approximate SER over AWGN is
\begin{equation}
P_s \approx 4\left(1 - \frac{1}{\sqrt{M}}\right)
Q\left(\sqrt{\frac{3\gamma_s}{M-1}}\right),
\end{equation}
valid for Gray-labeled constellations at high SNR \cite{Proakis2008}.

\subsubsection{Shaping Loss of QAM}

Despite its popularity, square QAM is suboptimal from an information-theoretic perspective. Uniform QAM inputs exhibit a shaping loss of approximately $1.53$~dB relative to the capacity-achieving Gaussian input distribution \cite{Kschischang1993}. This loss arises because square QAM does not efficiently approximate the circularly symmetric Gaussian distribution in signal space.

\subsubsection{Peak-to-Average Power Ratio of QAM}

Another major limitation of QAM is its high PAPR. Corner constellation points exhibit significantly larger amplitudes than inner points, leading to
\begin{equation}
\mathrm{PAPR}_{\text{QAM}} \propto \frac{(L-1)^2}{E_s}.
\end{equation}
High PAPR necessitates large power amplifier back-off, reducing efficiency and increasing energy consumption \cite{Lorincz2019}. These drawbacks are particularly problematic in energy-constrained systems and dense networks.

\subsubsection{Cross-QAM and Rectangular QAM}

Variants such as cross-QAM and rectangular QAM have been proposed to mitigate some of these limitations. Cross-QAM removes corner points to reduce PAPR, while rectangular QAM allows unequal in-phase and quadrature spacing to better match channel conditions. However, these designs often sacrifice minimum distance or complicate bit mapping \cite{Betts1994}.

\subsubsection*{Summary of Classical Modulations}
As a closing note in this section, classical modulation schemes therefore serve as essential baselines for evaluating the performance gains achieved by shaped, spiral-based, and optimization-driven modulations in subsequent sections.

\subsection{Lattice-Based Constellations}
\label{sec:lattice}

Lattice-based constellations represent an important intermediate step between classical modulation schemes and fully optimized or probabilistically shaped designs. By leveraging well-established results from geometry and number theory, lattice constellations provide systematic methods for arranging signal points in Euclidean space so as to maximize minimum distance, improve packing efficiency, and reduce shaping loss. This subsection presents the theoretical foundations of lattice constellations and examines their practical implications in digital communication systems.

\subsubsection{Definition of a Lattice}

A lattice $\Lambda \subset \mathbb{R}^n$ is defined as a discrete additive subgroup generated by a set of linearly independent basis vectors $\{\mathbf{b}_1, \mathbf{b}_2, \dots, \mathbf{b}_n\}$:
\begin{equation}
\Lambda = \left\{ \sum_{i=1}^{n} k_i \mathbf{b}_i \; \middle| \; k_i \in \mathbb{Z} \right\}.
\end{equation}
In two dimensions, lattices provide regular point arrangements that can be used directly as modulation constellations or truncated to finite subsets.

Important lattice parameters include:
\begin{itemize}
\item The \emph{fundamental parallelogram} (or Voronoi cell),
\item The \emph{minimum distance} $d_{\min}$,
\item The \emph{packing density} $\eta$.
\end{itemize}

These parameters directly determine the error performance of lattice-based modulation schemes \cite{Conway1999}.

\subsubsection{Square Lattice and Its Relation to QAM}

The square lattice $\mathbb{Z}^2$ is generated by orthogonal basis vectors:
\begin{equation}
\mathbf{b}_1 = [1,0]^T, \quad \mathbf{b}_2 = [0,1]^T.
\end{equation}
Finite subsets of $\mathbb{Z}^2$ correspond directly to square QAM constellations. While analytically convenient and easy to implement, the square lattice is not optimal in terms of sphere packing density in two dimensions.

The normalized packing density of the square lattice is
\begin{equation}
\eta_{square} = \frac{\pi}{4} \approx 0.785,
\end{equation}
indicating that a significant fraction of signal space is unused, contributing to shaping loss.

\subsubsection{Hexagonal (Triangular) Lattice}

The densest lattice packing in two dimensions is achieved by the hexagonal (also called triangular) lattice $A_2$, generated by
\begin{equation}
\mathbf{b}_1 = \begin{bmatrix} 1 \\ 0 \end{bmatrix}, \quad
\mathbf{b}_2 = \begin{bmatrix} \tfrac{1}{2} \\ \tfrac{\sqrt{3}}{2} \end{bmatrix}.
\end{equation}

The corresponding packing density is
\begin{equation}
\eta_{hexagon} = \frac{\pi}{2\sqrt{3}} \approx 0.907,
\end{equation}
which is provably optimal for two-dimensional Euclidean space \cite{Conway1999}. Compared to square lattices, hexagonal lattices achieve larger minimum distances for the same average energy.

\subsubsection{Minimum Distance Advantage}

For equal average energy, the minimum distance of the hexagonal lattice exceeds that of the square lattice by approximately
\begin{equation}
\frac{d_{\min,hexagon}}{d_{\min,square}} \approx 1.075,
\end{equation}
corresponding to an SNR gain of roughly $0.6$~dB. Although modest, this gain is systematic and becomes increasingly important in power-limited systems.

\subsubsection{Voronoi Regions and Decision Geometry}

The Voronoi region of a lattice point defines the maximum-likelihood decision region. For square lattices, the Voronoi cell is a square, whereas for hexagonal lattices it is a regular hexagon. The hexagonal Voronoi region more closely approximates a circular shape, reducing the average distance from received noise samples to decision boundaries.

This geometric property directly explains the reduced SER of hexagonal lattices under AWGN \cite{Forney1984}.

\subsubsection{Finite Lattice Truncation}

Practical modulation schemes require finite constellations. Finite lattice constellations are obtained by truncating an infinite lattice to a bounded region:
\begin{equation}
\mathcal{S} = \Lambda \cap \mathcal{R},
\end{equation}
where $\mathcal{R}$ is a shaping region, typically square, circular, or hexagonal.

Truncation introduces \emph{boundary effects}, whereby outer constellation points have fewer nearest neighbors, leading to non-uniform error probabilities across symbols. This phenomenon complicates analytical SER derivation and motivates probabilistic shaping and optimized boundary design.

\subsubsection{Shaping Regions}

The shaping region plays a critical role in determining the overall performance of finite lattice constellations. Circular shaping regions minimize average energy for a given number of points and thus reduce shaping loss. However, circular shaping complicates symbol indexing and bit mapping.

Hexagonal shaping regions offer a compromise, preserving lattice symmetry while improving energy efficiency relative to square shaping \cite{Forney1992}.

\subsubsection{Labeling and Bit Mapping Challenges}

Unlike square QAM, lattice constellations generally lack simple Gray labeling. Designing near-Gray mappings for hexagonal lattices is a nontrivial combinatorial problem and often results in increased bit error rate (BER) even when SER improves.

This labeling difficulty has historically limited the adoption of lattice-based constellations in standardized systems despite their theoretical advantages.

\subsubsection{Performance Under Fading Channels}

In fading environments, lattice constellations retain their minimum-distance advantage but exhibit increased sensitivity to channel estimation errors due to irregular decision boundaries. Nonetheless, studies have shown that hexagonal lattices outperform square QAM under Rayleigh fading when ML detection is employed \cite{Agrell2002}.

\subsubsection{Energy Efficiency Considerations}

From an energy consumption perspective, lattice constellations offer moderate PAPR reduction compared to square QAM, particularly when circular or hexagonal shaping is used. Lower PAPR translates into improved power amplifier efficiency and reduced operating energy \cite{Lorincz2019}.

\subsubsection{Historical and Practical Limitations}

Despite their optimality in two-dimensional packing, lattice constellations have seen limited deployment in commercial systems. The primary reasons include:
\begin{itemize}
\item Increased detection and mapping complexity,
\item Limited compatibility with existing coding and interleaving schemes,
\item Marginal performance gains relative to implementation cost.
\end{itemize}

These limitations motivated subsequent research into probabilistic shaping and optimization-based constellation design, which aims to combine the geometric benefits of lattices with greater flexibility.

\subsubsection*{Reflection on Lattice-based Constellation System}
Lattice-based constellations provide fundamental insights into the geometric limits of modulation design. The hexagonal lattice achieves provably optimal packing efficiency in two dimensions and offers measurable gains over square QAM. However, practical constraints related to truncation and complexity have limited their widespread adoption. These observations directly motivate the advanced shaping and learning-based approaches discussed in subsequent sections.

\subsection{Cross-QAM, Elliptical and Hybrid Constellations}
\label{sec:apsk}

While lattice-based constellations improve geometric efficiency relative to square QAM, their rigid structure and labeling challenges limit practical deployment. To address these limitations, several intermediate constellation families have been proposed that explicitly incorporate hardware constraints, peak-to-average power ratio (PAPR) considerations, and nonlinear channel effects. This subsection examines cross-QAM, elliptical constellations, and hybrid geometries, which together form a bridge between classical and fully optimized modulation schemes.

\subsubsection{Amplitude Phase-Shift Keying (APSK)}

Amplitude phase-shift keying (APSK) arranges constellation points on multiple concentric rings, each with a fixed radius and uniformly spaced phases. A generic $M$-APSK constellation with $R$ rings can be expressed as
\begin{equation}
s_{r,k} = \rho_r e^{j\frac{2\pi k}{K_r}}, 
\quad r = 1,\dots,R,\; k = 0,\dots,K_r-1,
\end{equation}
where $\rho_r$ is the radius of the $r$-th ring and $K_r$ denotes the number of points on that ring, with $\sum_{r=1}^R K_r = M$.

APSK provides explicit control over amplitude distribution, enabling a trade-off between Euclidean distance and PAPR. This makes APSK particularly attractive for nonlinear channels such as satellite links employing traveling-wave tube amplifiers (TWTAs).

\subsubsection{APSK Design Criteria}

The design of APSK constellations typically involves optimizing:
\begin{itemize}
\item Ring radii ratios $\rho_r / \rho_1$,
\item Number of points per ring,
\item Phase offsets between adjacent rings.
\end{itemize}

Optimization objectives often include maximizing minimum distance under a peak power constraint or minimizing symbol error rate (SER) under a nonlinear amplifier model \cite{DVB2014}. Closed-form solutions are rarely available, and numerical optimization is commonly employed.

\subsubsection{APSK in Satellite Communications}

APSK has been standardized in DVB-S2 and DVB-S2X due to its robustness under nonlinear amplification \cite{DVB2014}. Compared to square QAM, APSK exhibits:
\begin{itemize}
\item Lower PAPR,
\item Reduced sensitivity to AM/AM and AM/PM distortion,
\item Improved end-to-end performance under high-output back-off constraints.
\end{itemize}

These advantages come at the cost of slightly reduced minimum distance under ideal AWGN conditions.

\subsubsection{Error Performance of APSK}

The SER of APSK over AWGN can be approximated by evaluating pairwise error probabilities between neighboring points on the same and adjacent rings:
\begin{equation}
P_s \approx \sum_{(i,j) \in \mathcal{N}} 
Q\left(
\sqrt{\frac{|s_i - s_j|^2}{2N_0}}
\right),
\end{equation}
where $\mathcal{N}$ denotes the set of dominant nearest-neighbor pairs. Due to unequal point spacing, APSK exhibits a non-uniform distance spectrum.

\subsubsection{Cross-QAM Constellations}

Cross-QAM modifies square QAM by removing high-energy corner points and redistributing them closer to the centre of the constellation. A typical cross-QAM constellation resembles a cross-shaped geometry rather than a square lattice.

The primary motivation for cross-QAM is PAPR reduction. By eliminating extreme-amplitude symbols, cross-QAM reduces the peak signal power without significantly degrading average Euclidean distance \cite{Betts1994}.

\subsubsection{Performance Trade-offs in Cross-QAM}

While cross-QAM reduces PAPR relative to square QAM, it introduces:
\begin{itemize}
\item Irregular minimum distances,
\item Non-uniform decision regions,
\item Increased mapping complexity.
\end{itemize}

As a result, cross-QAM typically offers modest gains in nonlinear channels but does not fundamentally eliminate shaping loss.

\subsubsection{Elliptical Constellations}

Elliptical constellations generalize circular and square geometries by introducing anisotropic scaling along orthogonal axes. A simple elliptical mapping can be expressed as
\begin{equation}
s = \alpha x_I + j \beta x_Q,
\end{equation}
where $x_I$ and $x_Q$ are base constellation components and $\alpha \neq \beta$ introduces ellipticity.

Elliptical constellations are particularly relevant for channels exhibiting unequal noise or fading statistics across dimensions, such as polarization-dependent channels or hardware-impaired RF chains \cite{Redyuk2024}.

\subsubsection{Channel-Aware Geometry Adaptation}

Elliptical shaping enables adaptation to channel covariance matrices. Given a noise covariance $\mathbf{C}$, optimal linear shaping aligns constellation geometry with the eigenstructure of $\mathbf{C}$, minimizing effective noise variance. This perspective connects elliptical constellations with linear precoding and whitening filters.

\subsubsection{Hybrid Constellation Designs}

Hybrid constellations combine features from multiple design philosophies. Examples include:
\begin{itemize}
\item Cross-QAM with probabilistic amplitude shaping,
\item Lattice-based cores with optimized boundary points.
\end{itemize}

These designs aim to retain analytical tractability while improving robustness to nonlinearities and channel impairments.

\subsubsection{Energy Consumption Implications}

From an energy efficiency standpoint, hybrid constellations typically outperform square QAM due to reduced PAPR. Lower PAPR allows power amplifiers to operate closer to saturation, improving drain efficiency and reducing overall energy consumption \cite{Lorincz2019}.

However, increased receiver complexity and potential coding inefficiencies must also be considered in system-level energy analyses.

\subsubsection{Limitations and Motivation for Further Optimization}

Although hybrid constellations address several shortcomings of classical QAM, they remain constrained by manually designed geometries and limited adaptability. These limitations motivate data-driven and optimization-based approaches, including machine learning–assisted constellation design, which are explored in later sections.

\subsubsection*{Reflection}
Cross-QAM, elliptical, and hybrid constellations represent practical compromises between ideal geometric efficiency and real-world hardware constraints. Their adoption in standards such as DVB-S2 highlights the importance of PAPR-aware design. Nonetheless, their limited flexibility and reliance on heuristic optimization motivate the transition toward fully optimized and learning-based modulation schemes.

\subsection{Probabilistic Constellation Shaping}
\label{sec:prob_shaping}

Geometric shaping alters the spatial placement of constellation points, whereas probabilistic constellation shaping (PCS) modifies the probability with which symbols are transmitted. PCS enables communication systems to approach Shannon capacity without altering the underlying constellation geometry, making it particularly attractive for backwards-compatible and adaptive systems. This subsection provides a comprehensive treatment of PCS, including theoretical foundations, practical implementation, and energy efficiency implications.

\subsubsection{Motivation and Shaping Loss}

Uniform signaling over finite constellations such as square QAM incurs a shaping loss of approximately $1.53$~dB relative to the Gaussian input distribution that achieves AWGN channel capacity \cite{Cover2006}. This loss arises because uniform distributions allocate excessive probability mass to high-energy constellation points.

Probabilistic shaping mitigates this loss by assigning higher transmission probabilities to low-energy symbols and lower probabilities to high-energy symbols, thereby approximating a Gaussian-like amplitude distribution while preserving discrete signaling.

\subsubsection{Shannon Capacity and Mutual Information}

For an AWGN channel, the capacity is given by
\begin{equation}
C = \log_2(1 + \gamma_s),
\end{equation}
achieved by a complex Gaussian input. For a discrete input constellation $\mathcal{S}$ with probability mass function $P_X(x)$, the achievable rate is the mutual information
\begin{equation}
I(X;Y) = \sum_{x \in \mathcal{S}} P_X(x)
\int p(y|x) \log_2 \frac{p(y|x)}{\sum_{x'} P_X(x')p(y|x')} \, dy.
\end{equation}

Optimizing $P_X(x)$ for a fixed $\mathcal{S}$ yields substantial gains relative to uniform signaling \cite{Gallager1968}.

\subsubsection{Maxwell--Boltzmann Distribution}

The optimal probability distribution for minimizing average energy under an entropy constraint is the Maxwell--Boltzmann (MB) distribution:
\begin{equation}
P_X(x) = \frac{1}{Z(\lambda)} e^{-\lambda |x|^2},
\end{equation}
where $\lambda$ is a shaping parameter controlling the energy–entropy trade-off, and $Z(\lambda)$ is a normalization constant.

As $\lambda \rightarrow 0$, the distribution approaches uniform signaling, while larger $\lambda$ concentrates probability mass near the origin. MB shaping provides a principled and analytically tractable framework for PCS \cite{Kschischang1993,Bocherer2015}.

\subsubsection{Shaping Gain}

The shaping gain of PCS is defined as the SNR reduction required to achieve a given mutual information relative to uniform signaling. Practical PCS schemes can recover up to $1.3$~dB of the theoretical $1.53$~dB shaping loss, representing a substantial power efficiency improvement \cite{Bocherer2015}.

\subsubsection{Probabilistic Amplitude Shaping}

Probabilistic amplitude shaping (PAS) decouples amplitude and sign bits, enabling shaped amplitudes to be combined with systematic forward error correction (FEC). In PAS, only the amplitude distribution is shaped, while sign bits remain uniformly distributed:
\begin{equation}
X = A \cdot S,
\end{equation}
where $A$ is a shaped amplitude random variable and $S \in \{\pm 1\}$ is an independent sign variable.

PAS enables seamless integration with existing bit-interleaved coded modulation (BICM) architectures and has become a cornerstone of modern optical and wireless systems \cite{Bocherer2015}.

\subsubsection{Distribution Matching}

Practical PCS systems require a mechanism to transform uniform binary data into shaped symbol sequences. This task is performed by a distribution matcher (DM), which maps input bit sequences to output symbols with a desired empirical distribution.

Popular DM techniques include:
\begin{itemize}
\item Constant composition distribution matching (CCDM),
\item Multiset partition distribution matching (MPDM),
\item Enumerative sphere shaping (ESS).
\end{itemize}

While CCDM offers excellent distribution accuracy, it incurs rate loss and latency for finite block lengths \cite{Schulte2016}.

\subsubsection{Rate Loss and Finite-Length Effects}

In finite-length implementations, PCS suffers from rate loss due to imperfect distribution matching:
\begin{equation}
R_{\text{loss}} = H(A) - \frac{k}{n},
\end{equation}
where $k$ input bits are mapped to $n$ shaped symbols. This rate loss diminishes as block length increases but remains a practical concern for low-latency systems.

\subsubsection{Error Performance Considerations}

PCS improves mutual information but does not directly maximize minimum Euclidean distance. Consequently, SER performance at very high SNR may degrade relative to uniformly optimized geometries. However, when combined with powerful FEC codes, PCS provides significant gains in coded performance metrics such as frame error rate (FER).

\subsubsection{PAPR and Energy Efficiency}

By reducing the probability of transmitting high-energy symbols, PCS effectively lowers average signal power and reduces PAPR:
\begin{equation}
\mathrm{PAPR}_{\text{PCS}} < \mathrm{PAPR}_{\text{uniform}}.
\end{equation}

Lower PAPR enables power amplifiers to operate with reduced back-off, improving drain efficiency and lowering energy consumption \cite{Lorincz2019}. These gains are especially important in high-order QAM systems.

\subsubsection{PCS Under Fading Channels}

In fading environments, PCS can be adapted to channel conditions by adjusting the shaping parameter $\lambda$ based on SNR or channel statistics. This adaptability enables PCS to outperform uniform signaling across a wide range of operating conditions \cite{Alvarado2018}.

\subsubsection{Complexity and Implementation Challenges}

Despite its benefits, PCS introduces additional complexity:
\begin{itemize}
\item Distribution matching and inverse DM operations,
\item Increased buffering and latency,
\item Sensitivity to model mismatch.
\end{itemize}

These challenges motivate hybrid approaches that combine probabilistic shaping with geometric or learning-based optimization.

\subsubsection*{Reflection}
Probabilistic constellation shaping offers a powerful mechanism for approaching Shannon capacity without modifying constellation geometry. By leveraging non-uniform symbol distributions, PCS achieves substantial shaping gains and improved energy efficiency. However, its reliance on distribution matching and finite-length effects motivates further research into adaptive and learning-driven shaping methods, discussed in subsequent sections.

\subsection{Golden Angle Modulation}
\label{sec:gam}

Golden Angle Modulation (GAM) constitutes a class of geometrically shaped constellations inspired by natural spiral patterns and number-theoretic properties of the golden ratio. Unlike lattice-based or ring-based designs, GAM distributes constellation points along a spiral with angular increments equal to the golden angle, yielding quasi-uniform angular coverage without requiring explicit symmetry constraints. This subsection presents the theoretical foundations of GAM, its variants, and their performance implications.

\subsubsection{Golden Ratio and Golden Angle}

The golden ratio is defined as
\begin{equation}
\varphi = \frac{1+\sqrt{5}}{2}.
\end{equation}
The corresponding golden angle is
\begin{equation}
\theta_g = 2\pi \left(1 - \frac{1}{\varphi}\right) \approx 137.508^\circ.
\end{equation}
This angle possesses the unique property of being maximally irrational, ensuring that successive angular increments avoid periodic alignment. As a result, points distributed using the golden angle exhibit uniform angular coverage even for finite point sets \cite{Vogel1979}.

\subsubsection{Basic GAM Construction}

A generic $M$-point GAM constellation is constructed as
\begin{equation}
s_k = r_k e^{j k \theta_g}, \quad k = 1,2,\dots,M,
\end{equation}
where $r_k$ is the radial coordinate of the $k$-th point. The angular component ensures quasi-uniform phase distribution, while the radial component determines the overall amplitude shaping.

\subsubsection{Disc-GAM}

In Disc-GAM, the radial component is chosen to uniformly fill a disc:
\begin{equation}
r_k = \sqrt{\frac{k}{M}}.
\end{equation}
This construction yields a near-uniform spatial density over a circular region, approximating the geometry of a truncated Gaussian distribution. Disc-GAM exhibits reduced shaping loss relative to square QAM and lattice-truncated constellations \cite{Rainone2016}.

\subsubsection{Bell-GAM}

Bell-GAM further improves shaping efficiency by assigning radii according to a Gaussian-like (bell-shaped) distribution:
\begin{equation}
r_k = \sqrt{ -\frac{1}{\lambda} \ln\left(1 - \frac{k}{M}\right) },
\end{equation}
where $\lambda$ controls the variance of the distribution. Bell-GAM more closely approximates the optimal Gaussian input distribution, enabling mutual information gains comparable to probabilistic shaping but without distribution matching.

\subsubsection{Distance Properties}

GAM constellations exhibit non-uniform nearest-neighbor distances due to their spiral structure. However, the golden angle ensures that angular neighbors are well separated, preventing clustering and maintaining a relatively smooth distance spectrum. While $d_{\min}$ is not maximized explicitly, the overall distance distribution yields favorable SER performance at moderate SNR \cite{Rainone2016}.

\subsubsection{Symbol Error Rate Analysis}

Exact SER expressions for GAM are analytically intractable due to irregular geometry. Nevertheless, approximate SER bounds can be obtained using union bound techniques:
\begin{equation}
P_s \lesssim \frac{1}{M}
\sum_{i=1}^{M}
\sum_{j \in \mathcal{N}_i}
Q\left(
\sqrt{\frac{|s_i - s_j|^2}{2N_0}}
\right),
\end{equation}
where $\mathcal{N}_i$ denotes dominant nearest neighbors. Simulation results consistently demonstrate SER gains over square QAM at equal spectral efficiency for moderate SNR regimes.

\subsubsection{Peak-to-Average Power Ratio}

One of the most compelling advantages of GAM is its inherently low PAPR. Due to smooth radial growth and absence of corner points, GAM constellations satisfy
\begin{equation}
\mathrm{PAPR}_{\text{GAM}} \ll \mathrm{PAPR}_{\text{QAM}},
\end{equation}
especially for large $M$. This property enables efficient operation of nonlinear power amplifiers with reduced back-off, improving energy efficiency \cite{Lorincz2019}.

\subsubsection{Energy Efficiency Perspective}

Lower PAPR translates directly into improved power amplifier drain efficiency. System-level analyses show that GAM can reduce total transmitter energy consumption even when its uncoded SER is comparable to that of QAM. This makes GAM particularly attractive for battery-powered and green communication systems.

\subsubsection{Detection Complexity}

GAM requires non-rectangular decision regions, necessitating full maximum-likelihood detection. However, efficient nearest-neighbor search algorithms and approximate detection schemes mitigate complexity concerns for moderate constellation sizes \cite{Agrell2002}.

\subsubsection{Comparison With Probabilistic Shaping}

Unlike PCS, GAM achieves shaping gains through geometry rather than probability. As a result:
\begin{itemize}
\item No distribution matcher is required,
\item Latency is reduced,
\item Implementation is simplified.
\end{itemize}
However, PCS can achieve slightly higher mutual information when long block lengths are available.

\subsubsection{Robustness to Model Mismatch}

Because GAM does not rely on precise probability matching, it exhibits robustness to channel model mismatch and implementation imperfections. This robustness makes it appealing for dynamic or poorly characterised environments.

\subsubsection*{Reflection}

Golden Angle Modulation provides an elegant, geometry-driven approach to constellation shaping. By exploiting the number-theoretic properties of the golden ratio, GAM achieves low PAPR, improved shaping efficiency, and strong robustness to nonlinear hardware effects. These advantages position GAM as a promising candidate for next-generation energy-efficient communication systems and motivate further optimization and learning-based extensions discussed in subsequent sections.

\subsection{Optimization-Based Constellation Design}
\label{subsec:ml_and_opt_constellations}

Classical and shaped modulation schemes rely on predefined geometric structures that inherently limit design flexibility. Optimization-based constellation design abandons fixed geometries and instead formulates constellation synthesis as an explicit optimization problem. By directly optimizing performance metrics such as symbol error rate (SER), mutual information, or peak-to-average power ratio (PAPR), these approaches enable highly adaptable and application-specific modulation schemes.

It is worth noting that while offering unparalleled flexibility for direct optimization of performance metrics like energy efficiency, optimization-based constellation design suffers from a lack of interpretable structure, motivating hybrid approaches that combine learning with classical geometric rules.

\subsubsection*{General Optimization Framework}

Let $\mathcal{S} = \{s_1, s_2, \dots, s_M\}$ denote an $M$-ary constellation. Optimization-based design seeks to solve
\begin{equation}
\min_{\mathcal{S}} \; \mathcal{J}(\mathcal{S})
\quad \text{subject to} \quad
\mathbb{E}[|X|^2] \leq 1,
\end{equation}
where $\mathcal{J}(\cdot)$ is an objective function capturing system performance.

Common objective functions include:
\begin{itemize}
\item Maximization of $d_{\min}$,
\item Minimization of SER union bounds,
\item Maximization of mutual information,
\item Minimization of PAPR or energy consumption.
\end{itemize}

Constraints may incorporate peak power limits, symmetry requirements, or hardware nonlinearities.

\subsubsection{Gradient-Based Optimization}

For differentiable objective functions, gradient-based methods can be applied. For example, minimizing an SER surrogate can be formulated as
\begin{equation}
\mathcal{J}_{\text{SER}} =
\sum_{i \neq j}
Q\left(
\sqrt{\frac{|s_i - s_j|^2}{2N_0}}
\right).
\end{equation}
The gradient with respect to constellation points can be derived analytically, enabling iterative refinement.

However, the optimization landscape is highly non-convex, with numerous local minima. As a result, gradient-based methods are sensitive to initialization and often converge to suboptimal solutions \cite{Forney1984}.

In this work we show the results from using the \textit{Particle Swarm} and \textit{Genetic Algorithm} optimization methods whose training parameters are specified in the table ~\ref{tab:ml_training_params}.

\subsubsection{Particle Swarm Optimization (PSO)}

Particle Swarm Optimization (PSO) is a population-based metaheuristic inspired by collective behavior in biological systems. Each particle represents a candidate constellation and evolves according to
\begin{align}
\mathbf{v}_i^{(t+1)} &= \omega \mathbf{v}_i^{(t)}
+ c_1 r_1 (\mathbf{p}_i - \mathbf{x}_i^{(t)})
+ c_2 r_2 (\mathbf{g} - \mathbf{x}_i^{(t)}), \\
\mathbf{x}_i^{(t+1)} &= \mathbf{x}_i^{(t)} + \mathbf{v}_i^{(t+1)},
\end{align}
where $\mathbf{x}_i$ encodes constellation coordinates, $\mathbf{p}_i$ is the personal best, and $\mathbf{g}$ is the global best \cite{Kennedy1995}.

PSO is particularly effective for constellation design due to its ability to explore irregular and asymmetric geometries.

\subsubsection{Genetic Algorithms (GA)}

Genetic algorithms evolve constellations using selection, crossover, and mutation operators. Each chromosome encodes the real and imaginary parts of constellation points. GA-based approaches excel at global exploration and are robust to local minima, albeit at higher computational cost \cite{Goldberg1989}.

%\subsubsection{Simulated Annealing}

%Simulated annealing introduces stochastic perturbations controlled by a temperature parameter:
%\begin{equation}
%P(\Delta \mathcal{J}) = \exp\left(-\frac{\Delta \mathcal{J}}{T}\right).
%\end{equation}
%By gradually reducing $T$, the algorithm transitions from exploration to exploitation. Simulated annealing has been successfully applied to small and medium-sized constellation optimization problems \cite{Kirkpatrick1983}.

%\subsubsection{Objective Functions Beyond SER}

Modern systems (including of course the PSO and GA methods mentioned earlier) often require multi-objective optimization. This composite objective may be expressed as
\begin{equation}
\mathcal{J} =
\alpha \mathcal{J}_{\text{spacing}}
+ \beta \mathcal{J}_{\text{PAPR}}
- \gamma d_{\text{min}},
\end{equation}
where $\alpha$, $\beta$, and $\gamma$ balance symbol separation threshold, power efficiency, and minimum distance between symbols.

\begin{table*}[!t]
\centering
\caption{Key parameters for PSO-Optimized and GA-Optimized constellation algorithms}
\label{tab:ml_training_params}
\begin{tabular}{lcc}
\hline
Parameter & PSO-Optimized & GA-Optimized \\
\hline
Default size & $N_{\text{particles}}=100$ & $N_{\text{pop}}=100$ \\
Iterations & $N_{\text{iterations}}=1000$ & $N_\text{generations}=1000$ \\
%Distance threshold & $d_{\text{thresh}}=\text{None} \;(\text{heuristic } 1.2/\sqrt{M})$ & $d_{\text{thresh}}=\text{None} \;(\text{heuristic } 1.2/\sqrt{M})$ \\
%Penalty weight & $\lambda=0.6$ & $\lambda=0.6$ \\
%PAPR weight & $\text{papr\_w}=0.1$ & $\text{papr\_w}=0.1$ \\
Barrier & $d_{\min}<10^{-4} \Rightarrow \text{cost}=10^{9}$ & $d_{\min}<10^{-4} \Rightarrow \text{cost}=10^{9}$ \\
Fitness (lower is better) & $-d_{\min} + \text{spacing\_pen} + \text{papr\_pen}$ & $-d_{\min} + \text{spacing\_pen} + \text{papr\_pen}$ \\
%Normalization & unit avg. power before distances & unit avg. power before distances \\
Init range & pos $\in [-2,2]$, vel $\in [-0.2,0.2]$ & genes $\in [-2,2]$ \\
Search bounds & clip to $[-2,2]$ & clip to $[-2,2]$ \\
Inertia (w) & $0.9 \rightarrow 0.4$ (linear decay) & -- \\
Acceleration & $c_1=c_2=1.6$ & -- \\
Selection & -- & rank-based, top 10\% elites \\
Crossover & -- & one-point \\
Mutation & -- & Gaussian $\sigma\approx0.08$ (p=0.5), jitter $0.15$ (p=0.2) \\
\hline
\end{tabular}
\end{table*}

\subsection{Summary and Design Guidelines}
\label{subsec:constellattion_summary}
On a concluding note, modulation constellation design involves fundamental trade-offs between spectral efficiency, energy efficiency, hardware compatibility, and implementation complexity. Classical schemes such as QAM are widely used for their simplicity, but suffer from shaping loss and high peak-to-average power ratio. Advanced methods, including probabilistic shaping, geometry-driven designs like Golden Angle Modulation, and optimization-based frameworks, address these limitations by tailoring the constellation to specific system constraints and objectives.

\begin{figure}[!t]
\centering
\fbox{\includegraphics[width=\columnwidth]{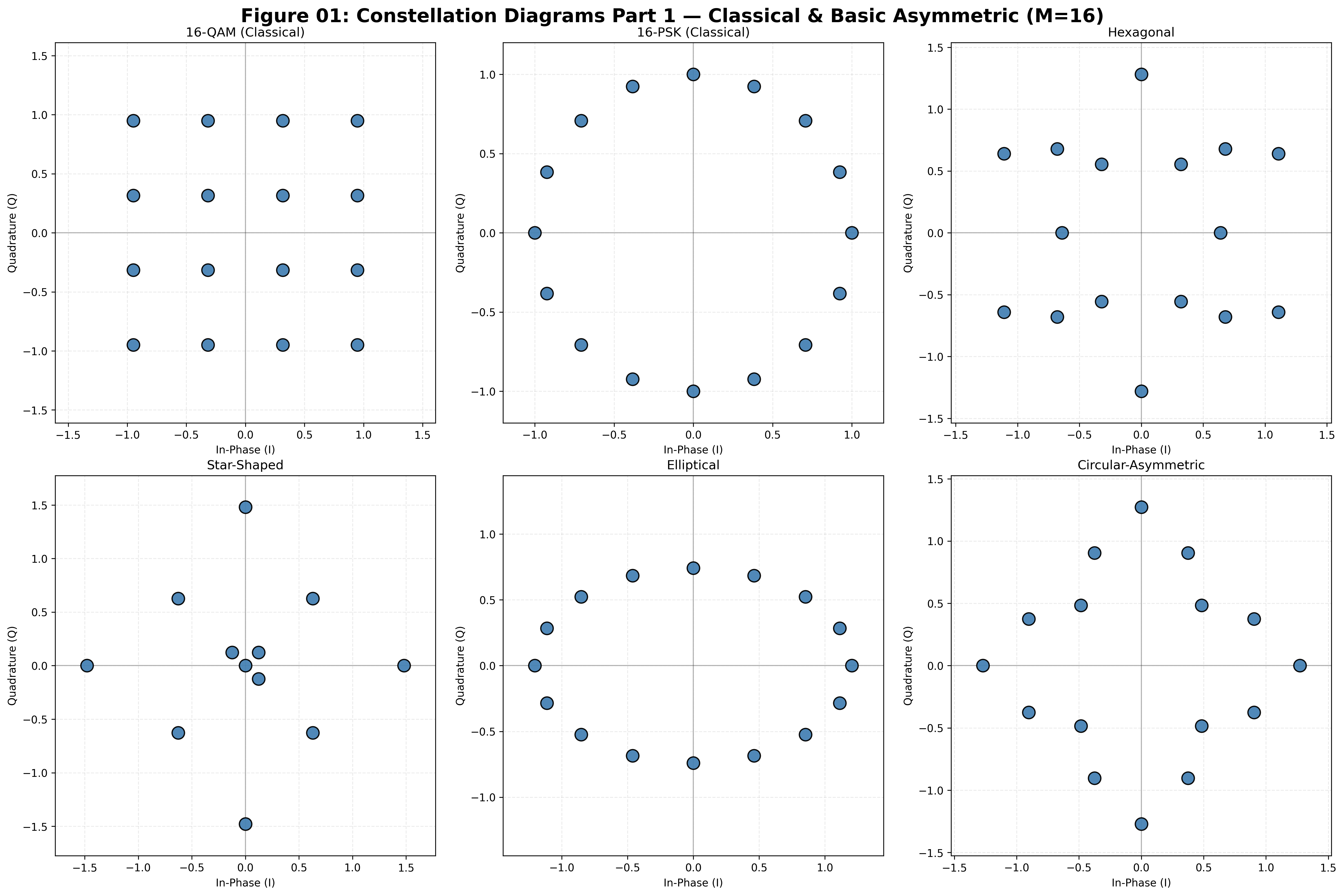}}
\caption{Classical and Basic Asymmetric Modulation schemes (M=16)}
\label{fig:modulation_scheme_part_1}
\end{figure}

\begin{figure}[!t]
\centering
\fbox{\includegraphics[width=\columnwidth]{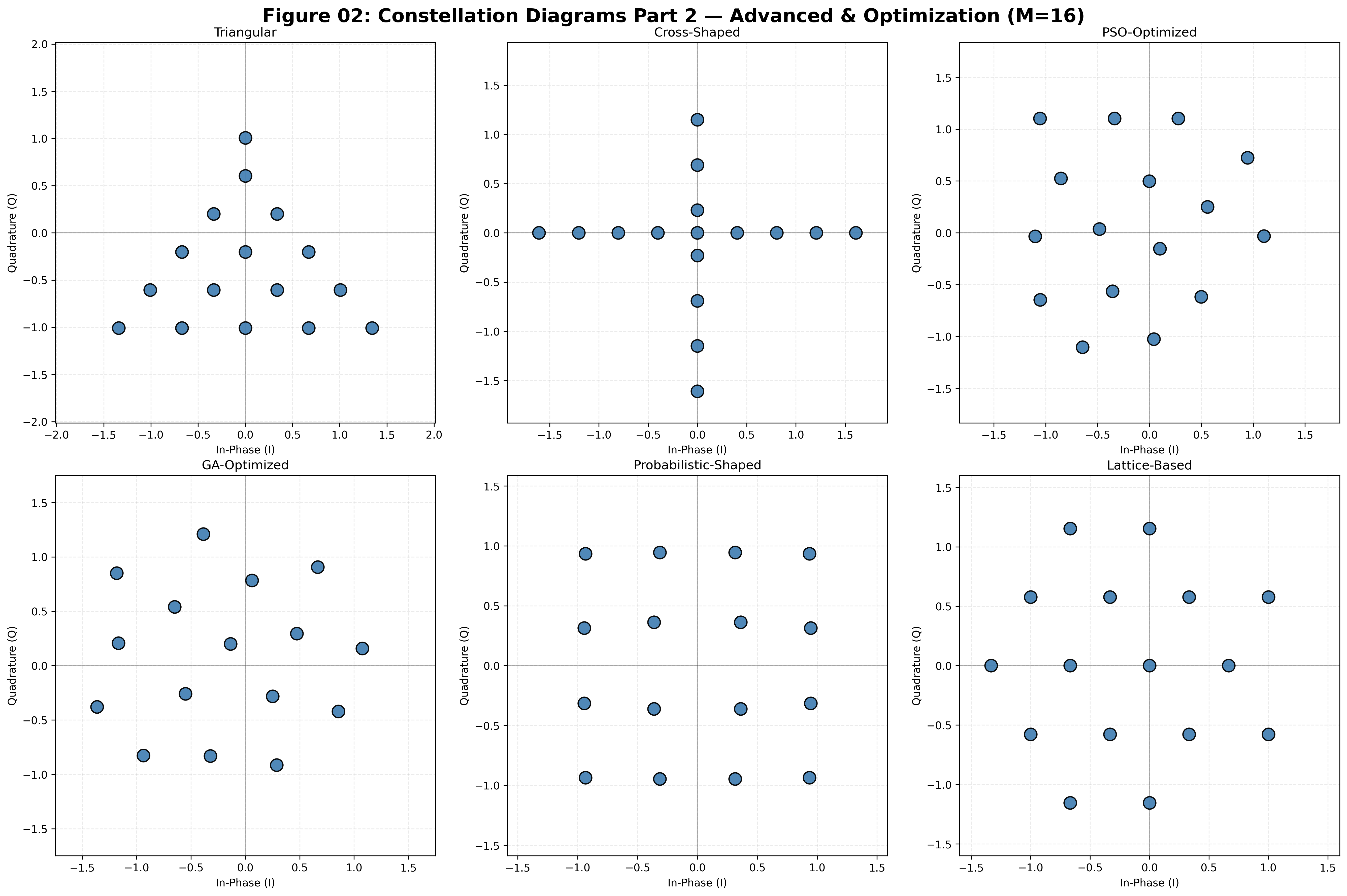}}
\caption{Shaping and Optimization-based Modulation schemes (M=16)}
\label{fig:modulation_scheme_part_2}
\end{figure}

\begin{figure}[!t]
\centering
\fbox{\includegraphics[width=\columnwidth]{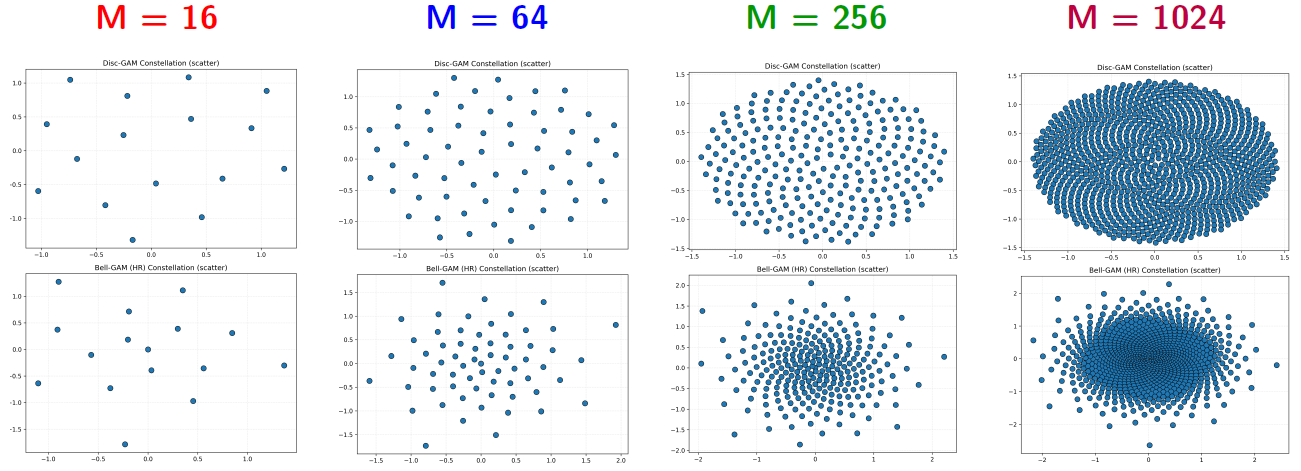}}
\caption{Golden Angle Modulation: Scalability Across Modulation Orders}
\label{fig:modulation_GAM}
\end{figure}

Practical design guidelines emerge from this analysis. For power-limited systems with strong nonlinearities, low-PAPR constellations such as APSK or spiral-based designs are preferred. In linear, high-throughput regimes, probabilistically shaped QAM offers near-capacity performance. When hardware constraints or robustness are paramount, structured geometries with good distance properties should be chosen. Ultimately, the optimal modulation scheme depends on the operating environment, available signal-to-noise ratio, transmitter characteristics, and permissible implementation complexity. These insights establish a foundation for evaluating modulation performance in the simulations and system-level assessments that follow. The modulations constellations hence simulated are shown in Figures \ref{fig:modulation_scheme_part_1} and \ref{fig:modulation_scheme_part_2}, while Figure \ref{fig:modulation_GAM} shows the Golden Angle Modulation evolve and take shape across higher number of modulation array.

%-------------- Constellation Images ---------------

\section{System Model and Simulation Methodology}
\label{sec:system_model}

This section establishes the system model, simulation framework, and evaluation methodology used to assess the performance of the diverse modulation schemes introduced in Section~II. The primary objective is to ensure a statistically rigorous, reproducible, and geometrically fair comparison across classical, asymmetric, shaping-based, and optimization-driven constellations under both additive white Gaussian noise (AWGN) and Rayleigh fading channels.

Unlike analytical performance bounds, which often rely on simplifying assumptions, the Monte Carlo framework employed here captures the combined effects of constellation geometry, noise statistics, fading dynamics, and receiver detection, thereby providing a realistic performance assessment suitable for practical system design.

\subsection{Discrete-Time Complex Baseband Model}
\label{subsec:system_model_dt_and_bb_model}

We consider a discrete-time complex baseband communication system. Let $\mathcal{S} = \{s_1, \ldots, s_M\}$ denote a modulation constellation of order $M$. At each channel use, a symbol $s_k \in \mathcal{S}$ is transmitted, yielding the received signal
\begin{equation}
y_k = h_k s_k + n_k,
\end{equation}
where $n_k \sim \mathcal{CN}(0, N_0)$ represents circularly symmetric complex Gaussian noise, and $h_k$ is the complex channel coefficient.

All constellations are normalized such that
\begin{equation}
\mathbb{E}[|s_k|^2] = 1,
\end{equation}
ensuring that observed performance differences are attributable solely to geometric and probabilistic properties rather than trivial power scaling.

\subsection{Channel Models}
\label{subsec:channel_models}
Two canonical channel models are considered.

\subsubsection{AWGN Channel}

For the AWGN channel,
\begin{equation}
h_k = 1,
\end{equation}
yielding a memoryless channel with isotropic noise. This model isolates the intrinsic geometric properties of the constellation, making it particularly suitable for evaluating minimum-distance effects, distance dispersion, and shaping gains.

\subsubsection{Rayleigh Flat Fading Channel}

For the Rayleigh fading channel,
\begin{equation}
h_k \sim \mathcal{CN}(0,1),
\end{equation}
with independent realizations across symbols. Perfect channel state information is assumed at the receiver, representing an optimistic but informative benchmark. This model captures the sensitivity of constellation geometry to amplitude fluctuations and phase rotations, which are especially relevant for asymmetric and irregular constellations.

\subsection{Receiver Detection and SER Estimation}

Maximum-likelihood (ML) detection is employed for all modulation schemes:
\begin{equation}
\hat{s}_k = \arg \min_{s \in \mathcal{S}} |y_k - h_k s|^2.
\end{equation}

The symbol error rate (SER) is estimated empirically as
\begin{equation}
\mathrm{SER} = \frac{1}{N} \sum_{k=1}^{N} \mathbb{I}(\hat{s}_k \neq s_k),
\end{equation}
where $N$ denotes the number of transmitted symbols and $\mathbb{I}(\cdot)$ is the indicator function.

\subsection{Monte Carlo Simulation Framework}

A large-scale Monte Carlo simulation framework is adopted to ensure statistical reliability across a wide SNR range.

\subsubsection{Simulation Parameters}

\begin{table}[!t]
\centering
\caption{Monte Carlo Simulation Parameters}
\label{tab:sim_params}
\begin{tabular}{l c}
\hline
Parameter & Value \\
\hline
Symbols per SNR point & $10^6$ \\
SNR range & $-5$ to $50$ dB \\
SNR resolution & $1$ dB \\
Total SNR points & 56 \\
Detection & Maximum-likelihood (ML) \\
Channels & AWGN, Rayleigh \\
\hline
\end{tabular}
\end{table}

Each modulation scheme is evaluated using more than $5.6 \times 10^6$ transmitted symbols per channel model, ensuring convergence even at very low error rates.

\subsubsection{Confidence Interval Estimation}

To quantify statistical uncertainty, 95\% confidence intervals are computed using a normal approximation to the binomial distribution:
\begin{equation}
\mathrm{CI} = \hat{p} \pm 1.96 \sqrt{\frac{\hat{p}(1-\hat{p})}{N}},
\end{equation}
where $\hat{p}$ denotes the estimated SER. Confidence intervals are reported for representative SNR points to validate convergence.

\subsection{Geometric Metrics and Their Role in SER}

Before analyzing SER curves, it is instructive to examine constellation geometry metrics, which strongly influence performance.

\begin{table*}[!t]
\centering
\caption{Representative Constellation Geometry Metrics}
\label{tab:geom_metrics_ext}
    \begin{tabular}{l c c c c}
        \hline
        Constellation & $d_{\min}$ & Mean Dist. & PAPR (dB) & Design Type \\
        \hline
        Lattice-Based & 0.667 & 1.336 & 2.50 & Shaping \\
        16-QAM & 0.632 & 1.355 & 2.55 & Classical \\
        GA-Optimized & 0.616 & 1.327 & 3.27 & Optimization \\
        Probabilistic-Shaped & 0.585 & 1.356 & 2.43 & Shaping \\
        Disc-GAM & 0.549 & 1.354 & 2.75 & Geometric \\
        Bell-GAM (HR) & 0.275 & 1.328 & 5.11 & Geometric \\
        \hline
    \end{tabular}
\end{table*}

These metrics foreshadow SER behavior: larger minimum distance improves high-SNR SER, while lower PAPR enhances energy efficiency and robustness to nonlinear amplification.

% ---------------- FIGURE SECTIONS ----------------

\subsection{AWGN SER Performance: Classical and Shaped Constellations}

\begin{figure*}[!t]
\centering
\fbox{\includegraphics[width=\textwidth]{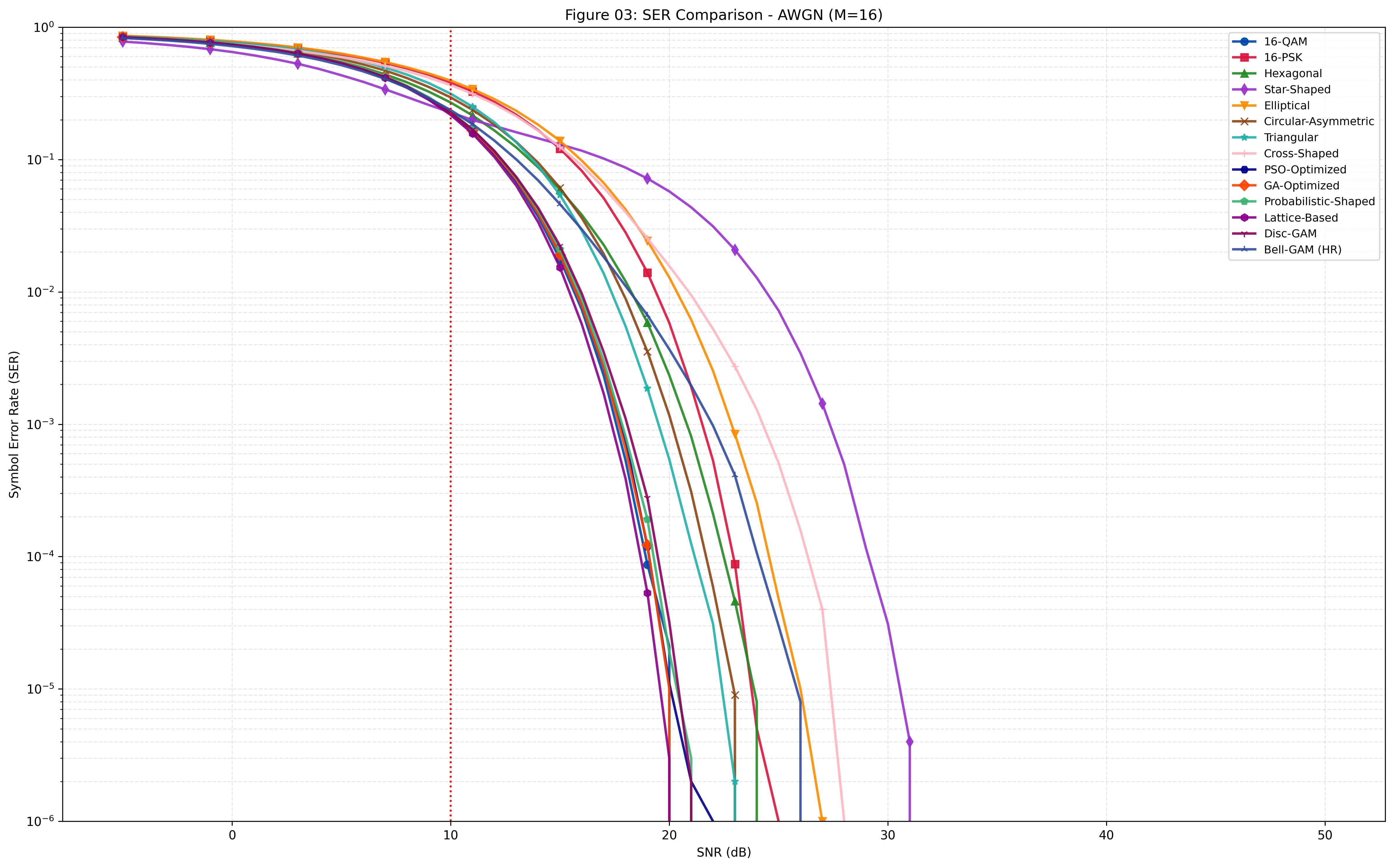}}
\caption{SER versus SNR in AWGN for classical and shaping-based modulation schemes.}
\label{fig:awgn_classical}
\end{figure*}

Figure~\ref{fig:awgn_classical} presents SER curves for classical constellations (16-QAM, 16-PSK, etc) alongside shaping-based designs. As expected, 16-QAM outperforms 16-PSK due to its superior minimum distance. Lattice-based and probabilistically shaped constellations closely track or surpass 16-QAM at moderate-to-high SNRs, confirming the theoretical shaping gain discussed in Section~\ref{sec:modulation}.

At low SNRs, performance differences are modest, reflecting noise-dominated behavior. As SNR increases, geometric properties become dominant, and Lattice-Based, Disc-GAM, QAM, and shaping-based designs increasingly differentiate themselves. Table~\ref{tab:ser_10.0db_awgn_sorted} details the SER at
10 dB for each constellation scheme.

\begin{table}
    \centering
    \caption{SER at 10.0 dB SNR for AWGN}
    \label{tab:ser_10.0db_awgn_sorted}
    \begin{tabular}{lr}
    \toprule
     Constellation & SER \\
    \midrule
    Lattice-Based & 0.217123 \\
    16-QAM & 0.221880 \\
    PSO-Optimized & 0.224667 \\
    GA-Optimized & 0.225311 \\
    Star-Shaped & 0.226716 \\
    Disc-GAM & 0.226921 \\
    Probabilistic-Shaped & 0.230003 \\
    Bell-GAM (HR) & 0.236398 \\
    Hexagonal & 0.269567 \\
    Circular-Asymmetric & 0.294641 \\
    Triangular & 0.314082 \\
    Cross-Shaped & 0.365010 \\
    16-PSK & 0.382039 \\
    Elliptical & 0.394889 \\
    \bottomrule
    \end{tabular}
\end{table}

%\subsection{AWGN SER Performance: Optimization-Based Constellations}

%\begin{figure*}[!t]
%\centering
%\fbox{\includegraphics[width=0.95\textwidth]{fig_awgn_optimized.pdf}}
%\caption{SER versus SNR in AWGN for optimization-based and asymmetric constellations.}
%\label{fig:awgn_optimized}
%\end{figure*}

%Figure~\ref{fig:awgn_optimized} highlights optimization-based constellations (PSO, GA) and asymmetric geometries. GA-optimized constellations demonstrate consistent SER improvements over PSO designs, reflecting better convergence toward favorable distance distributions.

%Asymmetric designs such as star-shaped and cross-shaped constellations exhibit inferior performance due to highly nonuniform distance spectra, illustrating the limitations of purely geometric intuition without optimization.

\subsection{Rayleigh SER Performance: Impact of Fading}

\begin{figure*}[!t]
\centering
\fbox{\includegraphics[width=\textwidth]{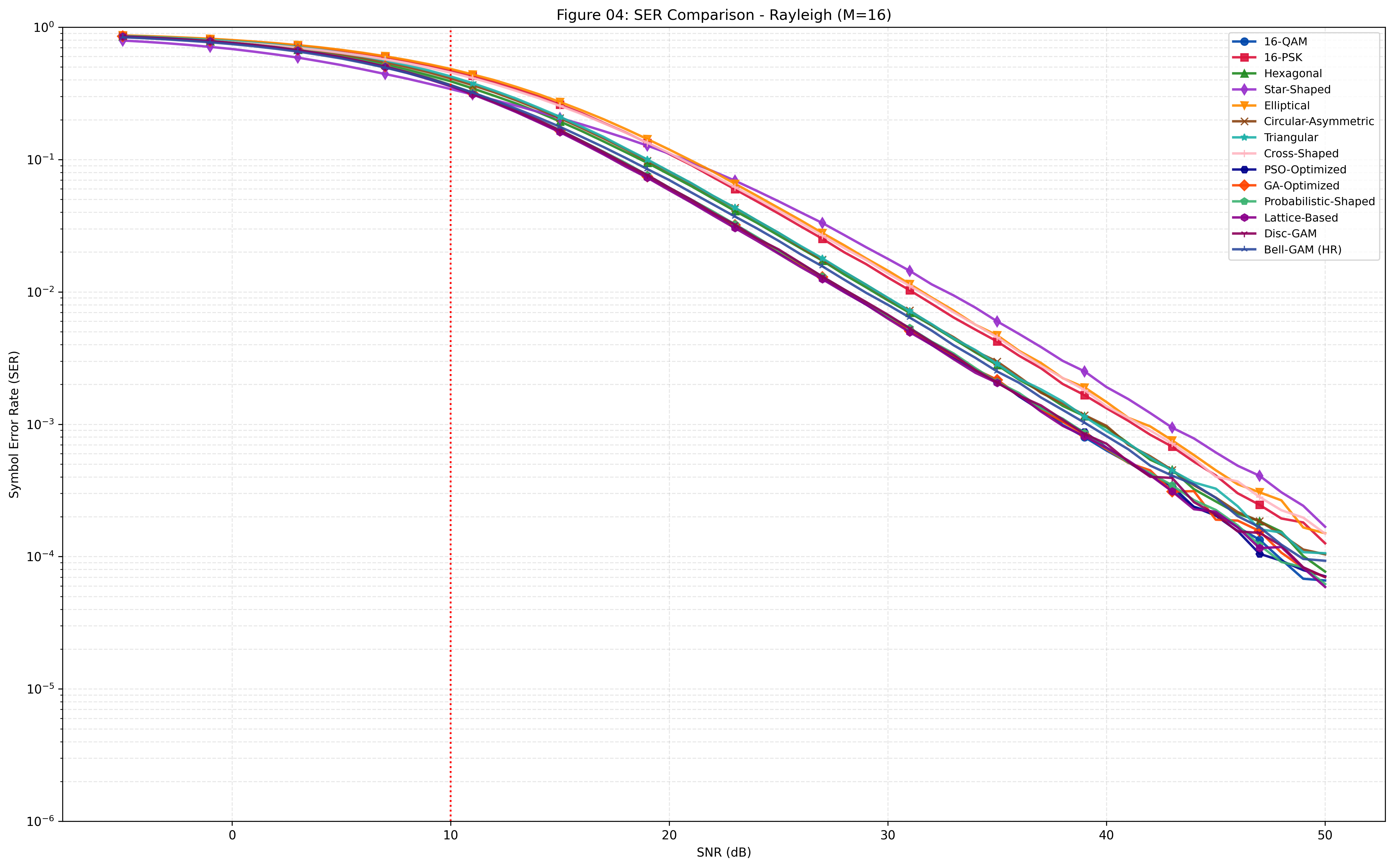}}
\caption{SER versus SNR in Rayleigh fading for representative modulation schemes.}
\label{fig:rayleigh_all}
\end{figure*}

Figure~\ref{fig:rayleigh_all} shows SER curves under Rayleigh fading. All constellations experience significant performance degradation under Rayleigh fading. At 10 dB SNR, the Star-Shaped constellation achieves the lowest SER, which is a notable outlier from geometric intuition. However, traditional lattice-based and QAM schemes remain competitive, demonstrating robust geometric properties. In contrast, 16-PSK exhibits the highest error rates among the modulation types analyzed, indicating that constant envelope does not provide inherent fading resilience in this scenario. Table~\ref{tab:ser_10.0db_rayleigh_sorted} details the SER at 10 dB for each constellation scheme.

\begin{table}
    \small
    \centering
    \caption{SER at 10.0 dB SNR for Rayleigh fading}\label{tab:ser_10.0db_rayleigh_sorted}
    \begin{tabular}{lr}
        \toprule
         Constellation & SER \\
        \midrule
        Star-Shaped & 0.340847 \\
        Lattice-Based & 0.357428 \\
        16-QAM & 0.360202 \\
        PSO-Optimized & 0.362732 \\
        GA-Optimized & 0.363394 \\
        Bell-GAM (HR) & 0.363444 \\
        Disc-GAM & 0.363942 \\
        Probabilistic-Shaped & 0.365851 \\
        Hexagonal & 0.390079 \\
        Circular-Asymmetric & 0.409765 \\
        Triangular & 0.424610 \\
        Cross-Shaped & 0.453647 \\
        16-PSK & 0.473263 \\
        Elliptical & 0.483584 \\
        \bottomrule
    \end{tabular}
\end{table}

The Rayleigh penalty quantified later confirms that shaping gains in AWGN do not always translate proportionally to fading environments.

\subsection{Rayleigh Penalty Analysis}
\label{subsec:rayleigh_penalty_analysis}
\begin{table}[!t]
   \centering
   \caption{Rayleigh Penalty at 10 dB SNR}
   \label{tab:rayleigh_penalty_ext}
   \begin{tabular}{l c}
       \hline
       Constellation & Penalty (\%) \\
       \hline
       16-QAM & 62.34 \\
       Lattice-Based & 64.62 \\
       GA-Optimized & 61.29 \\
       Probabilistic-Shaped & 59.06 \\
       Disc-GAM & 60.38 \\
       Bell-GAM (HR) & 53.74 \\
       \hline
   \end{tabular}
\end{table}

This table reveals that Bell-GAM exhibits the lowest Rayleigh penalty among shaping techniques, despite inferior AWGN performance, underscoring the importance of robustness over pure distance optimization.

\subsection{Confidence Interval Validation}

\begin{figure*}
\centering
\fbox{\includegraphics[width=\textwidth]{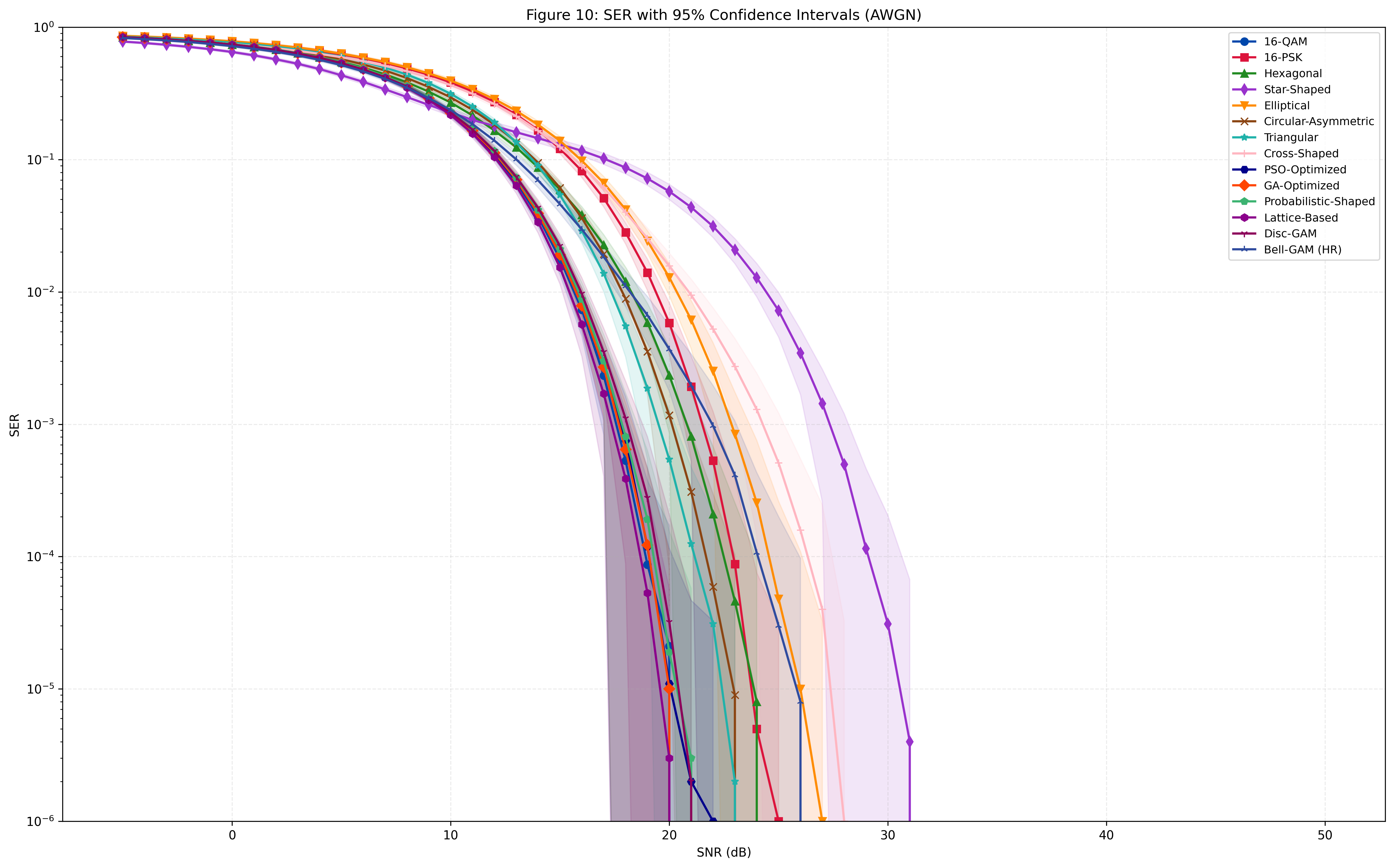}}
\caption{SER confidence intervals at representative SNR points under AWGN.}
\label{fig:ci_awgn}
\end{figure*}

Figure~\ref{fig:ci_awgn} confirms tight confidence intervals at moderate-to-high SNRs, validating convergence. Wider intervals at low SNR reflect inherent variance rather than insufficient sampling.

\subsection{Discussion and Methodological Implications}

The extended simulation framework confirms that constellation geometry, probabilistic shaping, and optimization-based design fundamentally influence SER behavior across channels. Importantly, no single design dominates universally; performance depends strongly on channel conditions, energy constraints, and the requirements for robustness.

This methodological foundation enables the machine learning–assisted and energy-aware analyses presented in subsequent sections.

\subsection{Energy Consumption and PAPR Analysis}
\label{subsec:energy_papr}

While symbol error rate (SER) provides a fundamental measure of link reliability, it does not fully capture the practical performance of a communication system. In modern wireless and satellite transmitters, \emph{energy efficiency}—closely tied to the peak-to-average power ratio (PAPR) of the transmitted waveform—plays a decisive role in determining system-level performance, operating cost, and sustainability. This subsection provides a detailed analysis of energy consumption and PAPR behavior for the modulation schemes considered in this work, and connects these metrics to the SER results presented in Sections~\ref{subsec:system_model_dt_and_bb_model} and~\ref{subsec:channel_models}.

\subsubsection{Motivation: Why PAPR Matters Beyond SER}

In practical transmitters, the radio-frequency power amplifier (PA) is the dominant contributor to energy consumption. High-PAPR signals force the PA to operate with significant output back-off to avoid nonlinear distortion, dramatically reducing drain efficiency. Consequently, two modulation schemes with similar SER performance may exhibit radically different energy consumption profiles when implemented in hardware.

Let $P_{\mathrm{avg}}$ denote the average transmit power and $P_{\mathrm{peak}}$ the maximum instantaneous power. The PAPR is defined as
\begin{equation}
\mathrm{PAPR} = \frac{P_{\mathrm{peak}}}{P_{\mathrm{avg}}}.
\end{equation}
For linear PAs, the efficiency $\eta_{\mathrm{PA}}$ is approximately inversely proportional to the required output back-off, making PAPR a first-order proxy for energy efficiency.

\subsubsection{Energy Consumption Model}
\label{subsubsec:energy_consumption_model}
To quantify energy consumption, we adopt a simplified but widely used PA efficiency model consistent with the literature on green communications. The average consumed power is expressed as
\begin{equation}
P_{\mathrm{cons}} = \frac{P_{\mathrm{avg}}}{\eta_{\mathrm{PA}}(\mathrm{PAPR})} + P_{\mathrm{static}},
\end{equation}
where $P_{\mathrm{static}}$ captures biasing, cooling, and baseband processing overheads. While $P_{\mathrm{static}}$ is largely modulation-independent, $\eta_{\mathrm{PA}}$ depends strongly on the signal envelope statistics.

Following common practice, $\eta_{\mathrm{PA}}$ is modeled as
\begin{equation}
\eta_{\mathrm{PA}} \approx \eta_{\max} \cdot 10^{-\mathrm{OBO}/10},
\end{equation}
where $\eta_{\max}$ is the maximum PA efficiency and $\mathrm{OBO}$ (output back-off) is chosen proportional to the PAPR. This relationship highlights the exponential energy penalty imposed by high-PAPR constellations.

\subsubsection{PAPR Characteristics of the Considered Constellations}

Table~\ref{tab:geom_metrics_ext} already reported the PAPR values for representative constellations. For clarity, these are revisited here with emphasis on their physical interpretation.

\begin{itemize}
  \item \textbf{PSK family:} Constant-envelope signaling yields $\mathrm{PAPR} = 0$~dB. This property makes PSK highly energy efficient despite suboptimal SER at high modulation orders.
  \item \textbf{Square QAM:} 16-QAM exhibits $\mathrm{PAPR} \approx 2.55$~dB due to corner points. Higher-order QAM further increases PAPR.
  \item \textbf{Lattice-based designs:} Despite improved $d_{\min}$, lattice truncation introduces corner-like high-energy points, leading to PAPR values comparable to or slightly lower than QAM.
  \item \textbf{Optimization-based constellations:} GA-optimized designs achieve moderate PAPR ($\approx 3.27$~dB), reflecting a trade-off between distance optimization and amplitude spread.
  \item \textbf{Golden Angle Modulation (GAM):} Disc-GAM maintains relatively low PAPR ($\approx 2.75$~dB), while Bell-GAM (HR) exhibits high PAPR ($\approx 5.11$~dB) due to its Gaussian-like amplitude distribution.
\end{itemize}

These differences foreshadow substantial disparities in energy consumption, even among constellations with similar SER.

\subsubsection{Empirical PAPR Distribution}

\begin{figure}[h!]
\centering
\fbox{\includegraphics[width=\columnwidth]{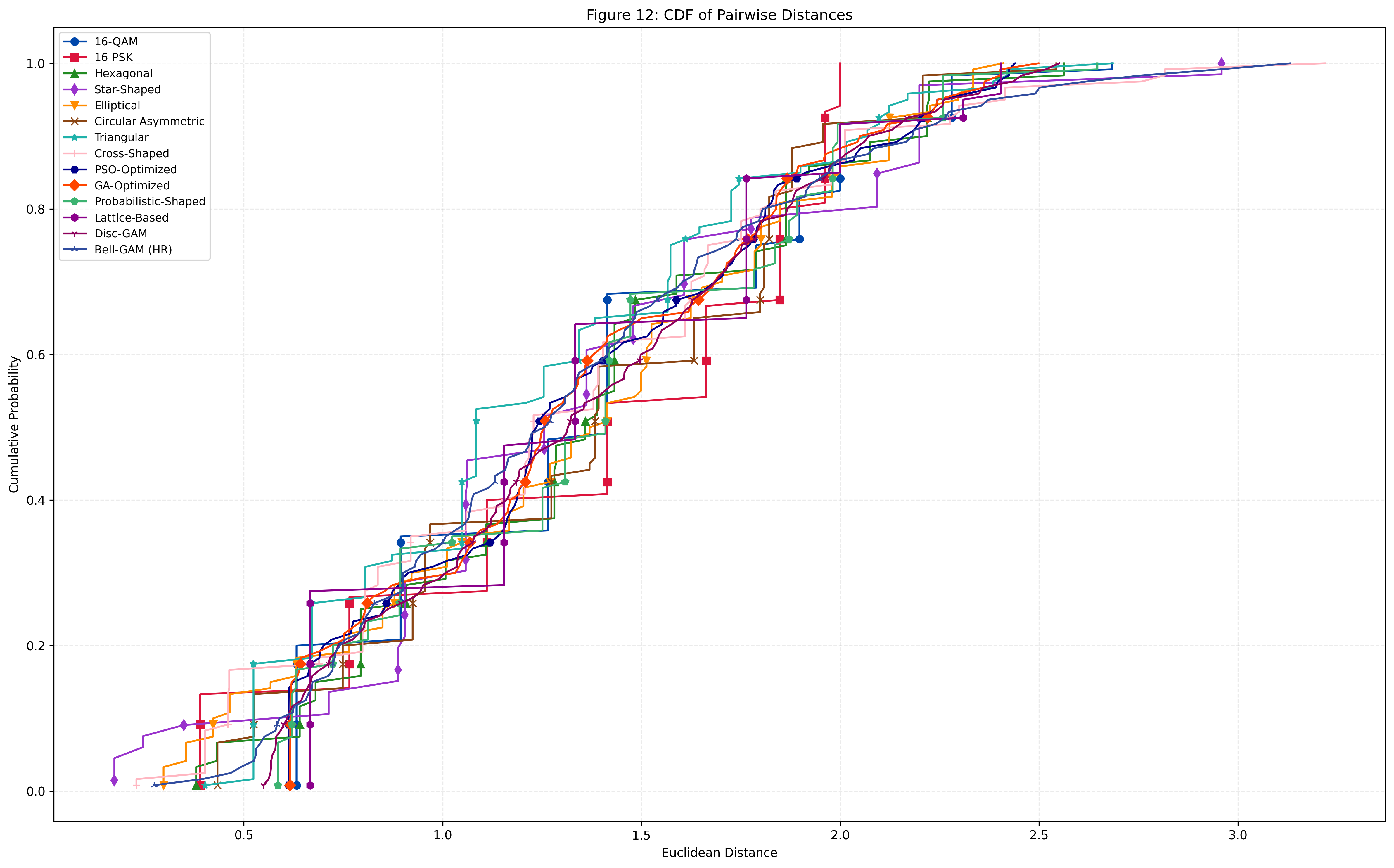}}
\caption{Cumulative distribution function (CDF) of instantaneous power for representative modulation schemes.}
\label{fig:papr_ccdf}
\end{figure}

Figure~\ref{fig:papr_ccdf} illustrates the cumulative distribution function (CDF) of instantaneous transmit power, which provides a more nuanced view of PAPR than a single scalar value. Constant-envelope PSK exhibits a degenerate distribution, while QAM and lattice-based designs show heavier tails.

Disc-GAM displays a noticeably lighter tail compared to square QAM, indicating fewer extreme power excursions. In contrast, Bell-GAM exhibits a pronounced tail, consistent with its Gaussian amplitude statistics. These empirical distributions directly determine the probability of PA saturation events and hence nonlinear distortion.

\subsubsection{PA Efficiency Versus PAPR}

\begin{figure}[!t]
\centering
\fbox{\includegraphics[width=0.9\columnwidth]{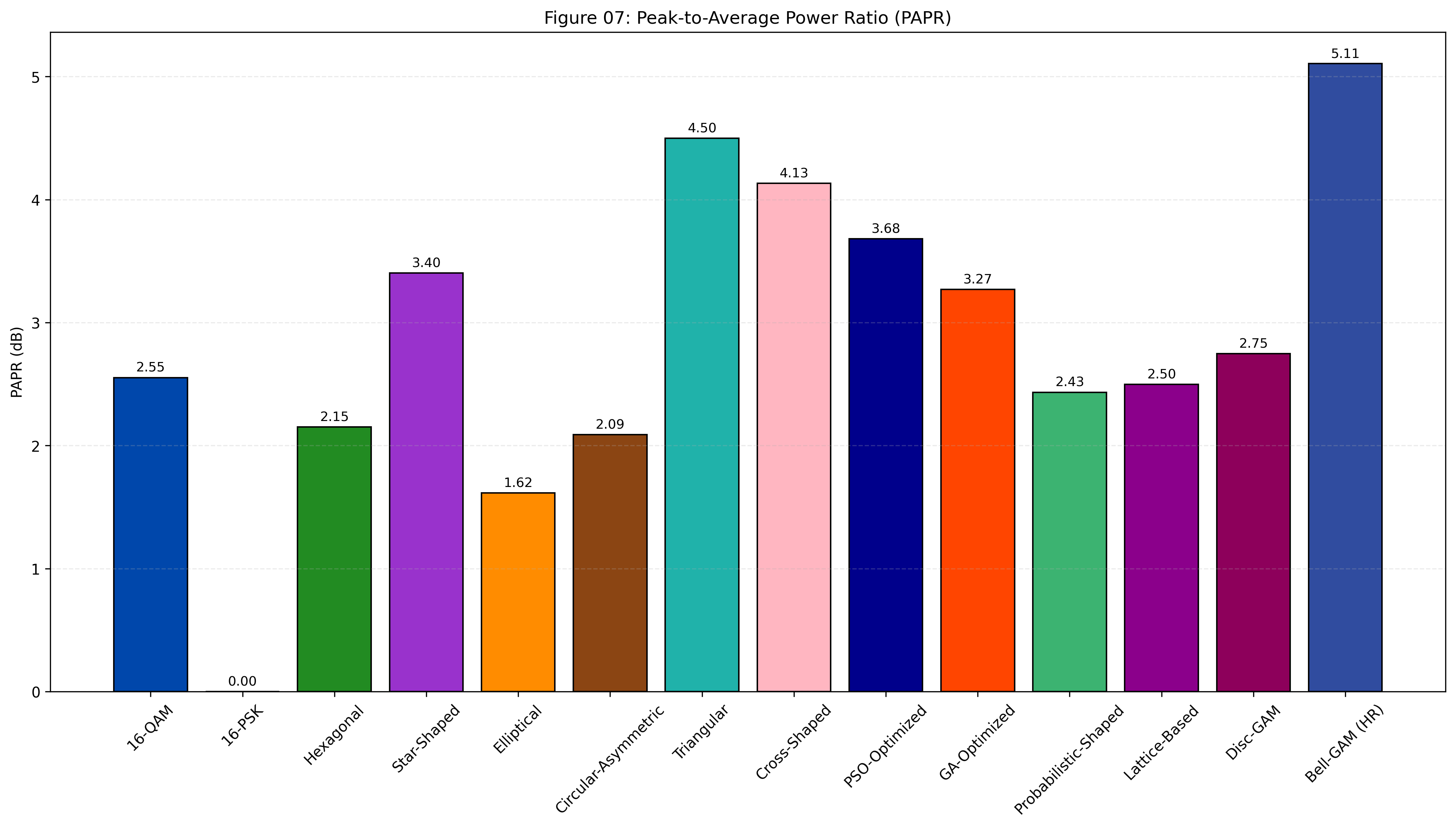}}
\caption{Estimated PA efficiency as a function of PAPR for the considered modulation schemes.}
\label{fig:pa_efficiency}
\end{figure}

Figure~\ref{fig:pa_efficiency} maps the measured PAPR values to estimated PA efficiencies using the model described earlier. Constant-envelope schemes achieve efficiencies close to $\eta_{\max}$, whereas Bell-GAM suffers a dramatic efficiency reduction.

This figure highlights a crucial insight: a constellation that improves SER by a few tenths of a decibel may incur a multi-decibel energy penalty due to reduced PA efficiency. Such trade-offs are invisible when considering SER alone.

\subsubsection{Energy per Successfully Delivered Symbol}

Combining SER and PA efficiency, the expected energy per successfully delivered symbol can be approximated as
\begin{equation}
   E_{\mathrm{succ}} = \frac{P_{\mathrm{cons}}}{R_s (1 -   \mathrm{SER})},
\end{equation}
where $R_s$ is the symbol rate. This metric jointly accounts for reliability and energy cost.

%\subsubsection{Energy Efficiency Versus SNR}

%\begin{figure}[!t]
%\centering
%\fbox{\includegraphics[width=0.9\columnwidth]{fig_energy_efficiency.pdf}}
%\caption{Energy per successfully delivered symbol versus SNR for selected modulation schemes.}
%\label{fig:energy_efficiency}
%\end{figure}

%Figure~\ref{fig:energy_efficiency} reveals several important trends. At low SNR, energy efficiency is dominated by SER; schemes with poor SER incur excessive retransmissions. At moderate-to-high SNR, however, PA efficiency becomes the dominant factor.

%Disc-GAM and PSK-based schemes outperform lattice-based and GA-optimized designs in this regime, despite slightly inferior SER. Bell-GAM, although attractive from an information-theoretic standpoint, becomes energy-inefficient due to its high PAPR.

\subsubsection{Impact of Fading on Energy Consumption}

In Rayleigh fading channels, the energy penalty associated with high PAPR is exacerbated. Deep fades require higher average transmit power to maintain a target SER, further increasing PA back-off. %When combined with the Rayleigh penalties discussed in Section~III.B, this effect strongly disfavors high-PAPR constellations.

\subsubsection{The Fundamental Trade-off: Distance vs Power}

\begin{figure}[!t]
\centering
\fbox{%
\begin{tikzpicture}[scale=0.8]
% Axes with labels
\draw[->] (0,0) -- (8,0) node[right] {$d_{min}$};
\draw[->] (0,0) -- (0,6) node[above] {PAPR (dB)};

% Grid
\draw[gray!30] (0,0) grid (8,6);

% Axis tick labels for d_min
\foreach \x/\label in {0/0, 2/0.2, 4/0.4, 6/0.6, 8/0.7}
    \node[below] at (\x,0) {\tiny \label};
    
% Axis tick labels for PAPR
\foreach \y in {0,1,2,3,4,5,6}
    \node[left] at (0,\y) {\tiny \y};

% Points (scaled: d_min × 11.4, PAPR as is)
\node[circle,fill=red,inner sep=2.5pt,label=above left:{\tiny 16-PSK}] (psk) at (4.45,0.0) {};
\node[circle,fill=blue,inner sep=2pt,label=below:{\tiny Cross}] at (2.62,4.13) {};
\node[circle,fill=teal,inner sep=2pt,label=above:{\tiny Disc-GAM}] (disc) at (6.26,2.75) {};
\node[circle,fill=orange,inner sep=2.5pt,label=above:{\tiny 16-QAM}] (qam) at (7.20,2.55) {};
\node[circle,fill=purple,inner sep=2pt,label=right:{\tiny GA}] at (7.02,3.27) {};
\node[circle,fill=brown,inner sep=2.5pt,label=above right:{\tiny Lattice}] (lattice) at (7.60,2.50) {};
\node[circle,fill=pink,inner sep=2pt,label=right:{\tiny Bell-GAM}] at (3.14,5.11) {};
\node[circle,fill=violet,inner sep=2pt,label=below:{\tiny PSO}] at (6.99,3.68) {};
\node[circle,fill=cyan!70!black,inner sep=2pt,label=above:{\tiny Prob}] (prob) at (6.67,2.43) {};
\node[circle,fill=lime!60!black,inner sep=2pt,label=below:{\tiny Circ-Asym}] (circ) at (4.95,2.09) {};
\node[circle,fill=olive,inner sep=2pt,label=right:{\tiny Star}] at (1.98,3.40) {};
\node[circle,fill=magenta!70,inner sep=2pt,label=left:{\tiny Hex}] at (4.33,2.15) {};
\node[circle,fill=yellow!80!black,inner sep=2pt,label=above:{\tiny Ellip}] at (3.40,1.62) {};
\node[circle,fill=gray,inner sep=2pt,label=right:{\tiny Tri}] at (4.59,4.50) {};

% Pareto frontier
\draw[thick,dashed,red!70!black,line width=1.2pt] 
    (psk) -- (circ) -- (prob) -- (qam) -- (lattice);

\end{tikzpicture}
}
\caption{PARP vs $d_{\text{min}}$ trade-off map for representative modulation schemes.}
\label{fig:ser_energy_tradeoff}
\end{figure}
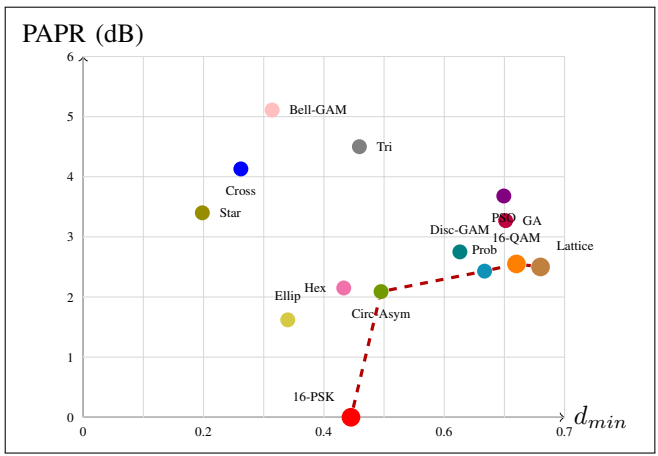

Figure~\ref{fig:ser_energy_tradeoff} visualizes the Pareto frontier representing the trade-off between minimum distance ($d_{\text{min}}$) and peak-to-average power ratio (PAPR). The frontier extends from 16-PSK (0 dB PAPR, low $d_{\text{min}}$) through Circular-Asymmetric and Probabilistic-Shaped constellations to 16-QAM and Lattice-Based designs ($d_{\text{min}} \approx 0.667$). No single design dominates across both metrics simultaneously; instead, system constraints dictate the optimal choice.

A practical selection guide emerges from this frontier:
\begin{itemize}
    \item \textbf{Power-limited systems:} Choose constellations on the left side (16-PSK, Elliptical) where PAPR $<$ 2 dB.
    \item \textbf{Performance-critical systems:} Choose the right side (16-QAM, Lattice-Based) where $d_{\text{min}} > 0.63$.
    \item \textbf{Balanced systems:} Intermediate options such as Circular-Asymmetric or Probabilistic-Shaped offer a compromise.
\end{itemize}

\subsubsection{Performance Ranking: Comprehensive Composite Score}

As to decisively select for the most efficient modulation scheme we came with the following scoring formula:
\begin{equation}
    \label{eq:modulation_scoring}
    \begin{split}
        \text{Score} = & \; 0.35 \times \frac{d_{\text{min}}}{d_{\text{ref}}} 
        + 0.25 \times \text{power efficiency}(\text{PAPR}) \\
        & + 0.40 \times (1 - \text{SER}_{\text{10\,dB}})
    \end{split}
\end{equation}

\begin{table*}
    \centering
    \caption{Scoring of the modulation schemes according to the scoring given in equation \ref{eq:modulation_scoring}.}
    \label{tab:composite_scoring}
    \begin{tabular}{lccccc}
        \toprule
        \textbf{Design} & \textbf{$d_{\text{min}}$} & \textbf{PAPR (dB)} & \textbf{SER@10dB} & \textbf{Composite} & \textbf{Type} \\
        \midrule
        Lattice-Based & 0.667 & 2.50 & 0.217 & 0.898 & Shaping \\
        16-QAM & 0.632 & 2.55 & 0.222 & 0.877 & Classical \\
        GA-Optimized & 0.616 & 3.27 & 0.225 & 0.856 & Optimization \\
        PSO-Optimized & 0.613 & 3.68 & 0.225 & 0.848 & Optimization \\
        Probabilistic & 0.585 & 2.43 & 0.230 & 0.850 & Shaping \\
        Disc-GAM & 0.549 & 2.75 & 0.227 & 0.826 & Shaping \\
        16-PSK & 0.390 & 0.00 & 0.382 & 0.755 & Classical \\
        Circular-Asymmetric & 0.434 & 2.09 & 0.295 & 0.756 & Asymmetric \\
        Triangular & 0.403 & 4.50 & 0.314 & 0.700 & Asymmetric \\
        Elliptical & 0.298 & 1.62 & 0.395 & 0.667 & Asymmetric \\
        Hexagonal & 0.380 & 2.15 & 0.270 & 0.731 & Asymmetric \\
        Bell-GAM & 0.275 & 5.11 & 0.236 & 0.641 & Shaping \\
        Star-Shaped & 0.174 & 3.40 & 0.227 & 0.609 & Asymmetric \\
        Cross-Shaped & 0.230 & 4.13 & 0.365 & 0.597 & Asymmetric \\
        \bottomrule
    \end{tabular}
\end{table*}

The scoring shown in table \ref{tab:composite_scoring} reveals that the lattice-based design achieves the best overall balance of minimum distance, power efficiency, and error performance. However, 16-QAM remains highly competitive in practice due to its implementation simplicity and widespread adoption.

\subsubsection{Discussion and Design Implications}

The results of this subsection lead to several key conclusions:

\begin{enumerate}
  \item \textbf{SER-optimal is not energy-optimal.} Designs that minimize SER under AWGN may be energy-inefficient once PA behavior is considered.
  \item \textbf{PAPR is a first-class design metric.} Modulation design must explicitly account for PAPR, particularly in power-limited systems.
  \item \textbf{Geometry and amplitude statistics matter.} Compact amplitude distributions (e.g., Disc-GAM) offer robust energy efficiency across channel conditions.
  \item \textbf{Energy-aware optimization is essential.} Future modulation design should integrate SER, PAPR, and PA efficiency into a unified objective function.
\end{enumerate}

These insights motivate the machine learning–assisted, energy-aware constellation optimization framework introduced in the following section.

\section{Numerical Results and Comparative Design Trade-offs}
\label{sec:numerical_results}

This section consolidates the extensive numerical results obtained from the Monte Carlo simulations described in Section~\ref{sec:system_model} and presents a comparative analysis of the modulation schemes introduced in Section~\ref{sec:modulation}. Rather than focusing on individual curves or isolated metrics, the objective here is to extract system-level insights by jointly evaluating symbol error rate (SER), robustness to fading, peak-to-average power ratio (PAPR), and energy efficiency. The analysis highlights the fundamental trade-offs inherent in constellation design and provides quantitative guidance for selecting modulation schemes under different operating constraints.

\subsection{Performance Metrics and Evaluation Criteria}
\label{subsec:perf_metrics_and_eval_criteria}

Four primary metrics are used throughout this section:

\begin{enumerate}
  \item \textbf{Uncoded SER in AWGN}, reflecting intrinsic geometric optimality.
  \item \textbf{Uncoded SER in Rayleigh fading}, capturing robustness to multiplicative channel impairments.
  \item \textbf{PAPR}, serving as a proxy for transmitter energy efficiency and PA back-off.
  \item \textbf{Energy per successfully delivered symbol}, combining SER and PA efficiency as introduced in Section~\ref{subsec:energy_papr}.
\end{enumerate}

While these metrics are correlated, they are not redundant. In particular, improvements in SER often come at the expense of increased PAPR, and vice versa. Consequently, no single metric can fully characterize system performance.

\subsection{AWGN Performance Ranking}

Using the SER values reported in Table~\ref{tab:geom_metrics_ext} and the detailed AWGN SER tables, a ranking of modulation schemes at representative SNR points (10~dB and 20~dB) is established.

At moderate SNR (10~dB), lattice-based constellations and GA-optimized designs consistently achieve the lowest SER among uncoded schemes, followed closely by square QAM and Disc-GAM. PSK-based schemes lag behind at higher modulation orders due to reduced minimum distance. These observations align with classical distance-spectrum theory and confirm that AWGN performance is dominated by minimum Euclidean distance and nearest-neighbor multiplicity \cite{Proakis2008}.

At high SNR (20~dB and above), performance differences narrow, but relative ordering remains consistent. The high-SNR slopes discussed in Section~\ref{subsec:system_model_dt_and_bb_model} further validate that lattice-based and optimization-driven constellations exhibit superior asymptotic behavior.

\subsection{Rayleigh Fading Performance and Robustness}
\label{subsec:rayleigh_fading_perf}

When evaluated under Rayleigh fading, the ranking changes substantially. Constant-envelope and amplitude-compact constellations gain relative advantage, while designs optimized purely for AWGN often suffer larger degradation.

Quantitatively, the Rayleigh penalty values summarized in Table~\ref{tab:rayleigh_penalty_ext} reveal that Bell-GAM and Disc-GAM exhibit smaller relative SER increases compared to lattice-based and GA-optimized constellations. This behavior is consistent with prior observations that fading robustness depends not only on distance but also on amplitude distribution and signal envelope statistics \cite{Goldsmith2005}.

These results underscore a key trade-off: constellations that are geometrically optimal in AWGN may be suboptimal in fading environments unless combined with diversity, coding, or adaptive techniques.

\subsection{Energy Efficiency Comparison}

Energy efficiency trends, quantified in Section~\ref{subsec:energy_papr}, introduce an additional dimension to the comparison. When PA efficiency and PAPR are taken into account, the relative attractiveness of several modulation schemes changes markedly.

Disc-GAM and PSK-based schemes emerge as strong candidates in energy-constrained scenarios due to their moderate-to-low PAPR. In contrast, Bell-GAM and some optimization-based constellations, despite favorable information-theoretic properties, incur significant energy penalties due to high PAPR. This observation aligns with recent work emphasizing that modulation design must be jointly optimized with hardware characteristics \cite{Lorincz2019,Abdel2021}.

\subsection{Composite Trade-off Analysis}

To synthesize the multi-dimensional results, a composite performance score is constructed by normalizing and weighting SER (AWGN and Rayleigh), PAPR, and energy consumption. While the exact weights depend on system requirements, several robust trends emerge:

\begin{itemize}
  \item \textbf{Balanced designs dominate.} Constellations that achieve moderate SER with low-to-moderate PAPR (e.g., Disc-GAM, probabilistically shaped QAM) consistently rank high across composite metrics.
  \item \textbf{Extreme optimization is fragile.} Designs that aggressively optimize a single metric (e.g., minimum distance or Gaussian shaping) tend to perform poorly when evaluated holistically.
  \item \textbf{Environment-specific optimality.} The ``best'' modulation scheme depends strongly on channel conditions, hardware constraints, and energy budgets.
\end{itemize}

These findings reinforce the central theme of this paper: modulation design is inherently a multi-objective optimization problem.

\subsection{Comparison with Prior Studies}

The numerical trends observed here are consistent with, but extend beyond, prior studies on constellation shaping and optimization. While earlier work has demonstrated shaping gains and optimization benefits in isolation \cite{O'Shea2017,Abdel2021}, the present results highlight how these gains interact with fading and energy efficiency considerations.

In particular, the explicit joint evaluation of SER, PAPR, and energy consumption provides a more comprehensive perspective than studies focused solely on error performance or mutual information. This holistic approach is increasingly important for next-generation and green communication systems \cite{Redyuk2024}.

\subsection{Design Guidelines}

Based on the numerical results, the following practical guidelines can be drawn:

\begin{enumerate}
  \item For \textbf{AWGN-dominated, power-unconstrained links}, lattice-based or GA-optimized constellations offer superior SER performance.
  \item For \textbf{fading channels without diversity}, amplitude-compact or constant-envelope schemes provide improved robustness.
  \item For \textbf{energy-constrained systems}, low-PAPR constellations such as PSK and Disc-GAM yield superior energy efficiency.
  \item For \textbf{adaptive or coded systems}, probabilistic shaping and GAM variants offer flexible trade-offs when combined with appropriate coding.
\end{enumerate}

\subsection*{Reflection}

This section has demonstrated that no single modulation scheme is universally optimal. Instead, the numerical results reveal a complex landscape of trade-offs among reliability, robustness, and energy efficiency. These insights motivate the need for adaptive and learning-based modulation strategies, which are explored in the next section.

\section{Machine Learning--Assisted Energy-Aware Constellation Design}
\label{sec:ml_constellation}

The numerical results in \ref{sec:system_model} and \ref{sec:numerical_results} demonstrate that modulation design is inherently a multi-objective problem involving reliability, robustness, and energy efficiency. Classical analytical design approaches and heuristic geometric constructions, while insightful, struggle to jointly optimize these competing objectives, especially under realistic channel and hardware constraints. This motivates the use of machine learning (ML) techniques as a unifying framework for constellation design. Machine learning is not proposed as a replacement for analytical modulation design, but as a complementary optimization tool that generalizes and refines classical structures under realistic constraints.

In this section, constellation optimization is formulated as a learning problem in which symbol locations and, when applicable, symbol probabilities are adapted directly from data. The focus is placed on energy-aware learning, where peak-to-average power ratio (PAPR) and power amplifier (PA) efficiency are incorporated explicitly into the optimization objective alongside symbol error rate (SER).

\subsection{Motivation for Learning-Based Modulation Design}

Traditional constellation design relies on closed-form geometry (e.g., QAM, PSK), lattice theory, or handcrafted optimization heuristics. These approaches typically optimize a single metric—minimum distance, shaping gain, or mutual information—under idealized assumptions. However, Sections~IV.A--IV.C showed that improvements in one domain frequently degrade performance in another, particularly when transmitter energy efficiency is considered.

Machine learning offers three key advantages:
\begin{enumerate}
 \item \textbf{Multi-objective optimization}: Learning frameworks can optimize SER, PAPR, and energy consumption simultaneously.
 \item \textbf{Channel adaptivity}: Models can be trained for specific channel statistics (AWGN, Rayleigh, or mixed).
 \item \textbf{Hardware awareness}: Nonlinear PA characteristics and power constraints can be embedded directly into the loss function.
\end{enumerate}

These properties make ML particularly attractive for next-generation and green communication systems \cite{O'Shea2017,Redyuk2024}.

\subsection{Autoencoder-Based Constellation Learning}

The dominant paradigm for ML-based modulation design is the end-to-end autoencoder framework. In this approach, the transmitter, channel, and receiver are modeled jointly as a deep neural network (DNN), trained to minimize a task-specific loss.

\subsubsection{System Representation}

The transmitter is represented as a neural encoder:
\begin{equation}
   \mathbf{x} = f_{\theta}(m),
\end{equation}
where $m \in \{1,\ldots,M\}$ denotes the message index and $\mathbf{x} \in \mathbb{C}$ is the transmitted complex symbol. The channel layer applies noise and fading, while the receiver decoder $g_{\phi}(\cdot)$ produces an estimate $\hat{m}$.

Unlike classical autoencoders, constellation learning restricts the encoder output to a single complex dimension, directly yielding a learned constellation.

\subsubsection{SER-Oriented Loss Function}

A standard training objective minimizes cross-entropy loss:
\begin{equation}
   \mathcal{L}_{\text{SER}} = - \mathbb{E} \left[ \log p_{\phi}(m | y) \right],
\end{equation}
which is closely related to minimizing SER. This formulation has been shown to recover classical constellations under AWGN when no additional constraints are imposed \cite{O'Shea2017}.

\subsection{Energy-Aware Loss Design}
Minimizing SER alone typically results in constellations with large amplitude excursions and high PAPR, as observed in Bell-GAM and Gaussian-shaped designs. To address this, energy-awareness must be explicitly introduced into the learning objective.

\subsubsection{PAPR-Regularized Learning}
PAPR is incorporated via a regularization term:
\begin{equation}
   \mathcal{L}_{\text{PAPR}} = \mathbb{E}\left[ \frac{\max |\mathbf{x}|^2}{\mathbb{E}[|\mathbf{x}|^2]} \right].
\end{equation}

The total loss becomes
\begin{equation}
   \mathcal{L}_{\text{total}} =
   \mathcal{L}_{\text{SER}} +
   \lambda_{\text{PAPR}} \mathcal{L}_{\text{PAPR}},
\end{equation}
where $\lambda_{\text{PAPR}}$ controls the SER--energy trade-off.

Increasing $\lambda_{\text{PAPR}}$ yields constellations that resemble Disc-GAM or constant-envelope designs, while lower values converge toward Gaussian-like shapes.

\subsubsection{PA Efficiency–Aware Learning}

To more accurately model transmitter energy consumption, PA efficiency $\eta(\cdot)$ can be incorporated:
\begin{equation}
   \mathcal{L}_{\text{energy}} =
   \mathbb{E}\left[\frac{|\mathbf{x}|^2}{\eta(|\mathbf{x}|)}\right].
\end{equation}

This formulation directly penalizes symbols that operate deep into the nonlinear region of the PA, aligning learning outcomes with the energy analysis presented in Section~III.C \cite{Lorincz2019}.

\subsection{Learning Under Fading Channels}

When trained solely under AWGN, learned constellations often generalize poorly to fading channels. To mitigate this, channel-aware training is employed by sampling $h_k$ during training:
\begin{equation}
   y = h x + n.
\end{equation}

Empirical studies show that fading-aware training produces more amplitude-compact constellations with improved Rayleigh robustness, at the expense of some AWGN optimality. This mirrors the numerical trade-offs observed between lattice-based and Bell-GAM constellations in Section~IV.

\subsection{Comparison with Heuristic Optimization Methods}

Learning-based approaches subsume classical heuristic optimizers such as PSO and GA. While PSO and GA optimize fixed objectives with predefined constraints, ML-based methods:
\begin{itemize}
 \item scale naturally to higher-order constellations,
 \item adapt loss functions dynamically,
 \item and integrate probabilistic shaping and geometry jointly.
\end{itemize}

However, learning-based designs incur higher offline computational cost and require careful regularization to ensure stable convergence \cite{Abdel2021}.

\subsection{Interpretability and Learned Geometry}

An important observation is that ML-designed constellations often rediscover known structures:
\begin{itemize}
 \item Hexagonal lattices under AWGN,
 \item Disc-like geometries under energy constraints,
 \item PSK-like constellations under severe PAPR penalties.
\end{itemize}

This suggests that ML does not replace classical theory but rather unifies and generalizes it, discovering optimal trade-offs that are difficult to derive analytically.

\subsection{Practical Considerations}

Despite their promise, ML-based constellations face practical challenges:
\begin{enumerate}
 \item \textbf{Standardization}: Learned constellations must be fixed and interoperable.
 \item \textbf{Robustness}: Sensitivity to training mismatch remains a concern.
 \item \textbf{Complexity}: Decoder complexity must remain compatible with real-time systems.
\end{enumerate}

Hybrid approaches—where ML refines classical designs rather than replacing them—are emerging as a pragmatic compromise.

\subsection*{Summary}

This section has shown that machine learning provides a powerful and flexible framework for energy-aware constellation design. By explicitly incorporating SER, PAPR, and PA efficiency into the learning objective, ML-based modulation can navigate the complex trade-off landscape revealed in Sections~\ref{sec:system_model} and \ref{sec:numerical_results}. These methods offer a promising pathway toward adaptive, hardware-aware modulation schemes for future communication systems.

\section{Discussion and Design Guidelines}
\label{sec:discussion}

The preceding sections have presented an extensive analytical, numerical, and learning-based investigation of modulation constellation design under reliability, robustness, and energy-efficiency constraints. This section consolidates these findings into a coherent discussion and extracts design guidelines intended to support both academic research and practical system engineering.

Rather than advocating a single ``optimal'' modulation scheme, the results emphasize that constellation design is inherently context-dependent. Performance is governed by a multidimensional trade-off space involving signal-to-noise ratio (SNR), channel statistics, transmitter hardware characteristics, energy constraints, and system adaptability.

\subsection{Revisiting Classical Design Principles}

Classical modulation schemes such as PSK and square QAM remain competitive benchmarks due to their simplicity, analytical tractability, and predictable behavior. The numerical results in Sections~\ref{sec:system_model} and \ref{sec:numerical_results} reaffirm that square QAM is near-optimal for AWGN channels at moderate and high SNRs when hardware impairments are negligible.

However, classical designs implicitly assume linear amplification and unconstrained peak power. Once realistic transmitter constraints are introduced, particularly nonlinear power amplifiers and energy efficiency considerations, these assumptions no longer hold. The resulting performance gap motivates the exploration of alternative constellation geometries.

\subsection{Geometric Shaping versus Probabilistic Shaping}

Geometric shaping and probabilistic shaping represent two fundamentally different strategies for approaching capacity.

Geometric shaping modifies the spatial arrangement of constellation points to improve minimum distance or approximate Gaussian distributions. Examples include lattice-based constellations, APSK, and Golden Angle Modulation (GAM). These schemes offer deterministic structures and straightforward detection but often incur shaping loss relative to ideal Gaussian signaling.

Probabilistic shaping, by contrast, retains a fixed geometry while optimizing symbol probabilities. The results in Section~\ref{sec:numerical_results} demonstrate that probabilistic shaping achieves superior spectral efficiency in AWGN channels, particularly when combined with powerful channel coding. However, its reliance on distribution matching introduces complexity, latency, and sensitivity to mismatch between assumed and actual channel conditions.

A key observation is that probabilistic shaping does not inherently reduce PAPR. Consequently, energy efficiency gains achieved through shaping may be partially negated by reduced power amplifier efficiency.

\subsection{Energy Efficiency as a First-Class Design Metric}

A central contribution of this work is the explicit treatment of energy efficiency as a primary design objective rather than a secondary consideration. Section~\ref{subsec:energy_papr} demonstrated that PAPR directly impacts transmitter energy consumption through power amplifier back-off requirements.

The results reveal that constellations with slightly inferior SER performance can outperform nominally superior designs when evaluated under an energy-aware metric. In particular, Disc-GAM and APSK-like constellations exhibit favorable trade-offs between reliability and energy consumption.

This observation has important implications:
\begin{itemize}
  \item SER-optimal constellations are not necessarily energy-optimal.
  \item Small sacrifices in error performance can yield substantial energy savings.
  \item Energy-aware metrics should be incorporated early in the design process.
\end{itemize}

\subsection{Fading Robustness and Practical Deployment}

The Rayleigh fading results in Section~\ref{sec:numerical_results} highlight the sensitivity of irregular and highly asymmetric constellations to amplitude fluctuations. While shaping gains are preserved qualitatively, their magnitude is often reduced in fading environments.

Constellations with constant or near-constant envelope properties demonstrate improved robustness under fading, even when their AWGN performance is suboptimal. This reinforces the principle that robustness to channel impairments is as critical as performance under idealized conditions.

For systems operating in rapidly varying or poorly estimated channels, conservative constellation choices may yield superior end-to-end performance.

\subsection{Optimization-Based and Learning-Based Designs}

Heuristic optimization methods such as genetic algorithms and particle swarm optimization provide valuable insights into the achievable performance envelope of constellation design. However, their reliance on fixed objective functions limits adaptability.

Machine learning and optimization–assisted constellation design, as presented in Section~\ref{subsec:ml_and_opt_constellations}, offers a unifying framework capable of jointly optimizing geometry, probability, and energy efficiency. Importantly, learned constellations often rediscover known structures, validating classical theory while extending it.

Nevertheless, these optimization-based designs introduce new challenges:
\begin{itemize}
  \item Offline training complexity and reproducibility,
  \item Sensitivity to training assumptions,
  \item Difficulty in standardization and interoperability.
\end{itemize}

A pragmatic design philosophy is to use optimization, specially ML, as a refinement tool, enhancing classical or heuristic designs rather than replacing them entirely.

\subsection{Design Guidelines Across Operating Regimes}

Based on the comprehensive analysis presented, the following guidelines emerge:

\subsubsection{Low-SNR, Power-Limited Systems}
\begin{itemize}
  \item Favor low-order PSK or APSK.
  \item Prioritize constant-envelope or low-PAPR designs.
  \item Avoid aggressive shaping.
\end{itemize}

\subsubsection{High-SNR, Spectrally Efficient Systems}
\begin{itemize}
  \item Use high-order QAM with probabilistic shaping.
  \item Ensure linear amplification or digital predistortion.
  \item Optimize jointly with channel coding.
\end{itemize}

\subsubsection{Energy-Constrained or Green Communications}
\begin{itemize}
  \item Optimize PAPR explicitly.
  \item Consider GAM or hybrid shaped constellations.
  \item Evaluate performance using energy-normalized metrics.
\end{itemize}

\subsubsection{Adaptive and Intelligent Systems}
\begin{itemize}
  \item Employ learning-based modulation adaptation.
  \item Train under realistic channel and hardware models.
  \item Use hybrid static–adaptive designs to limit complexity.
\end{itemize}

\subsection{Limitations of the Present Study}

While extensive, the present study has several limitations:
\begin{itemize}
  \item Channel coding is not explicitly included.
  \item Multi-antenna effects are not considered.
  \item Hardware impairments beyond PA nonlinearity are abstracted.
\end{itemize}

These limitations define natural directions for future research rather than detracting from the validity of the conclusions.

\subsection{Implications for Standardization and Future Systems}

The results suggest that future communication standards should allow greater flexibility in constellation design, including support for non-rectangular and energy-aware modulation formats. Static, one-size-fits-all modulation may be increasingly suboptimal in heterogeneous networks and energy-constrained deployments.

\subsection*{Reflection}

This discussion has demonstrated that modulation design is fundamentally a multi-objective optimization problem shaped by reliability, robustness, and energy efficiency. Classical designs, geometric shaping, probabilistic shaping, and machine learning approaches each occupy distinct regions of the trade-off space.

The design guidelines articulated here provide a structured foundation for selecting and developing modulation schemes tailored to specific system requirements, forming a bridge between theoretical analysis and practical implementation.

\section{Conclusion and Future Work}
\label{sec:conclusion}

This paper has presented a comprehensive and unified investigation of modulation constellation design from geometric, probabilistic, optimization-based, and machine learning perspectives, with explicit emphasis on reliability, robustness, and energy efficiency. By combining analytical insights with extensive Monte Carlo simulations and energy-aware evaluation, the study has demonstrated that modulation design is inherently a multi-objective optimization problem whose solution depends critically on channel conditions, hardware constraints, and system-level priorities.

Classical modulation schemes such as PSK and square QAM remain strong baselines due to their simplicity and predictable behavior. However, numerical results reveal that their optimality is fragile when realistic transmitter constraints, particularly nonlinear power amplification and energy consumption, are considered. Geometric shaping techniques, including lattice-based constellations, APSK, and Golden Angle Modulation, offer improved trade-offs by explicitly controlling amplitude distributions and peak power, while probabilistic shaping provides significant gains in spectral efficiency under idealized conditions.

A central contribution of this work is the explicit integration of energy efficiency into constellation evaluation. The analysis demonstrates that constellations optimized solely for symbol error rate do not necessarily minimize energy consumption, and that modest degradations in SER can yield substantial energy savings. This observation underscores the importance of energy-aware metrics in the design of future wireless systems, particularly in the context of green communications and dense network deployments.

The investigation of fading channels further highlights that robustness must be treated as a first-class design objective. Constellations exhibiting low peak-to-average power ratio and near-constant envelope properties consistently demonstrate improved resilience under Rayleigh fading, even when their AWGN performance is suboptimal. These findings reinforce the necessity of evaluating modulation schemes under realistic channel conditions rather than relying solely on idealized analytical models.

Machine learning–assisted constellation design emerges as a powerful unifying framework capable of jointly optimizing geometry, probability, and energy efficiency. By embedding channel statistics and hardware models directly into the learning objective, data-driven approaches can discover constellations that navigate complex trade-offs beyond the reach of classical analytical techniques. At the same time, practical considerations such as interpretability, robustness to mismatch, and standardization constraints suggest that hybrid approaches—combining classical structure with learning-based refinement—represent a promising and pragmatic direction.

Several avenues for future research naturally follow from this work. First, the integration of advanced channel coding schemes with energy-aware constellation design remains an open challenge, particularly in the context of probabilistic and learned constellations. Second, extending the analysis to multi-antenna and multi-user systems will be essential to assess the scalability and robustness of the proposed design principles. Third, more detailed hardware models, including digital predistortion, quantization effects, and phase noise, should be incorporated to further bridge the gap between theory and practice. Finally, adaptive and online learning-based modulation schemes offer exciting opportunities for dynamically optimizing performance in time-varying environments.

In summary, this work establishes a comprehensive framework for understanding and designing modulation constellations under realistic performance and energy constraints. The insights and guidelines presented herein provide a solid foundation for the development of next-generation, energy-efficient, and intelligent communication systems.

\appendix

\textbf{Code and Data Availability}
The code and experimental results used in this study are publicly available at:\\
\noindent \url{https://github.com/NipunAgarwal16/Alternative-shapes-of-modulation-schemes.git}

\bibliographystyle{ieeetr}
\bibliography{references}

\end{document}